# From Classical Trapping to Curvature-Induced Bolus Formation in Peristaltic Wave–Driven Flows

Lucija Josipovic,[1] Melissa Reyes Munoz,[1] Fernando Mendez Estrada,[1] Hui Wan,[2] Fangjun Shu,[1] and Yanxing Wang [1 †]

[1]*Department of Mechanical and Aerospace Engineering, New Mexico State University, Las Cruces, NM 88003, USA*
[2]*Department of Mechanical and Aerospace Engineering, University of Colorado, Colorado Springs, CO 80918, USA*

Bolus formation is one of the most distinctive features of peristaltic transport, yet the mechanisms governing its development remain incompletely understood. In this work, axisymmetric peristaltic-wave-driven flows under zero-pressure-gradient conditions are investigated using a high-fidelity lattice Boltzmann framework over a broad range of wave amplitudes, wavelengths, and Reynolds numbers spanning the Stokes limit to (Re=100). In the Stokes limit, bolus formation is governed by the classical trapping mechanism and exhibits a remarkable wavelength-normalized similarity, with both bolus morphology and transport characteristics approaching common asymptotic behaviors as the wavelength increases. As fluid inertia becomes significant, this similarity progressively breaks down and wavelength-dependent trapping behavior emerges. In the short-wavelength regime, the pronounced geometric bending of the wave reorganizes the local flow structure and produces a distinct off-axis bolus associated with a curvature-induced trapping mechanism. The influence of this mechanism expands toward longer wavelengths with increasing Reynolds number and occupies an increasingly large portion of the parameter space considered in the present study. The results reveal that bolus formation cannot be described by a single universal mechanism. Instead, the observed trapping behavior arises from the competition between classical trapping and curvature-induced trapping, whose relative importance is controlled by wavelength, wave amplitude, and Reynolds number. These findings establish a unified physical framework for understanding bolus formation and transport in peristaltic-wave-driven flows.

## I. INTRODUCTION

Peristaltic-wave-driven transport is a ubiquitous mechanism in physiological and engineering systems, where propagating waves of wall deformation drive fluid motion in fluid-filled distensible tubes. Early investigations were motivated primarily by physiological pumping problems and focused on understanding how wave motion generates pressure rise and volumetric flow rate. Foundational theoretical studies by Burns and Parkes [1], Hanin [2], Fung and Yih [3], Shapiro et al. [4], and Yin and Fung [5], and related investigations established the hydrodynamic basis of peristaltic pumping, while experiments by Latham [6], Weinberg [7], and Eckstein [8] provided validation of the emerging theory. These developments were synthesized by Jaffrin and Shapiro [9], who consolidated the understanding of peristaltic pumping and transport and established a unified framework for interpreting the effects of wave amplitude, wavelength, Reynolds number, pressure rise, and flow rate.

A major advance was achieved by Shapiro et al. [4], who developed a long-wavelength, low-Reynolds-number theory for peristaltic flow with arbitrary wave amplitude. Beyond establishing quantitative

† Contact author: yxwang@nmsu.edu

pressure–flow relationships, they examined particle trajectories and streamline topology within the travelling wave and identified two fundamental transport phenomena: reflux and trapping. Reflux demonstrated that fluid particles may undergo net motion opposite to the overall pumping direction, whereas trapping corresponds to the formation of a closed recirculating fluid region, or bolus, that travels with the wave. These concepts fundamentally changed the interpretation of peristaltic transport by revealing that material transport cannot be inferred solely from global pumping characteristics and established the foundation for subsequent studies of bolus transport and trapping dynamics.

Subsequent efforts focused on understanding how finite wavelength and fluid inertia modify the classical long-wavelength theory. Jaffrin [10] extended the analysis to finite wavelengths and moderate Reynolds numbers and demonstrated that streamline curvature and inertia substantially influence pumping, reflux, and trapping. Numerical investigations by Brown and Hung [11], Takabatake and Ayukawa [12], Pozrikidis [13], Takabatake et al. [14], and Xiao and Damodaran [15] progressively relaxed the assumptions underlying earlier asymptotic theories and systematically examined finite-amplitude, finite-wavelength, and finite-inertia peristaltic flows in both two-dimensional channels and axisymmetric tubes. These studies quantified the effects of Reynolds number, wavelength, and wave amplitude on transport, trapping, reflux, pressure distribution, and wall shear stress, and established much of the current understanding of finite-wavelength and finite-inertia peristaltic transport.

More recently, peristaltic-flow research has expanded toward increasingly complex physical conditions and application-oriented problems. Representative examples include finite-length transport, multilayer and non-Newtonian fluids, asymmetric geometries, fluid–structure interaction, microfluidic mixing, intestinal transport, bacterial population dynamics, gastric mixing, and drug-delivery systems [16-26].

Despite these extensive developments, the physical mechanisms governing bolus formation and evolution remain incompletely understood. Previous studies established the existence of trapping and systematically documented its dependence on Reynolds number, wavelength, and wave amplitude [4,10-15]. However, these investigations primarily focused on characterizing how trapping changes with flow parameters rather than explaining why trapping occurs or identifying the physical processes responsible for its formation [10,12-15]. The origins of several observations reported in the literature remain unclear, including the strong dependence of trapping on wavelength and inertia and the variations in bolus size and topology observed under different flow conditions. In particular, it remains unknown whether the trapping phenomena observed across different parameter regimes originate from a common physical mechanism or arise from distinct transport processes.

In the present work, we perform a systematic investigation of axisymmetric peristaltic-wave-driven flows over a broad range of wavelengths, amplitudes, and Reynolds numbers. By examining flow topology, transport characteristics, and bolus dynamics, we focus on the mechanisms responsible for bolus formation. The analysis reveals a previously unrecognized similarity behavior in the Stokes limit, establishes the physical mechanism governing classical trapping, and demonstrates that trapping is not a single phenomenon but can emerge through distinct physical processes. In particular, we identify a curvature-induced trapping mechanism that arises at short wavelengths and represents an alternative trapping pathway from the classical trapping mechanism.

## II. PHYSICAL MODEL

As illustrated in Fig. 1, we consider an incompressible fluid confined within a circular tube whose wall is described by an infinite train of sinusoidal waves. The tube has a mean radius $R$, while the wave amplitude and wavelength are denoted by $A$ and $\lambda$, respectively. In the laboratory frame of reference, every point on the tube wall undergoes periodic radial motion described by a sinusoidal function, causing the wall deformation to propagate downstream (toward the positive axial direction) at a constant wave speed $c_w$. The instantaneous tube radius is given by

$$h(x,t) = R + Asin[2\pi\,(x - c_w t)/\lambda + \phi] \tag{2.1}$$

where $h(x,t)$ denotes the local tube radius at axial location $x$ and time $t$, and $\phi$ is the initial phase angle of the travelling wave. The radial velocity of the wall follows directly from Eq. (2.1),

$$\frac{\partial h(x,t)}{\partial t} = -\frac{2\pi c_w A}{\lambda} cos[2\pi\,(x - c_w t)/\lambda + \phi] \tag{2.2}$$

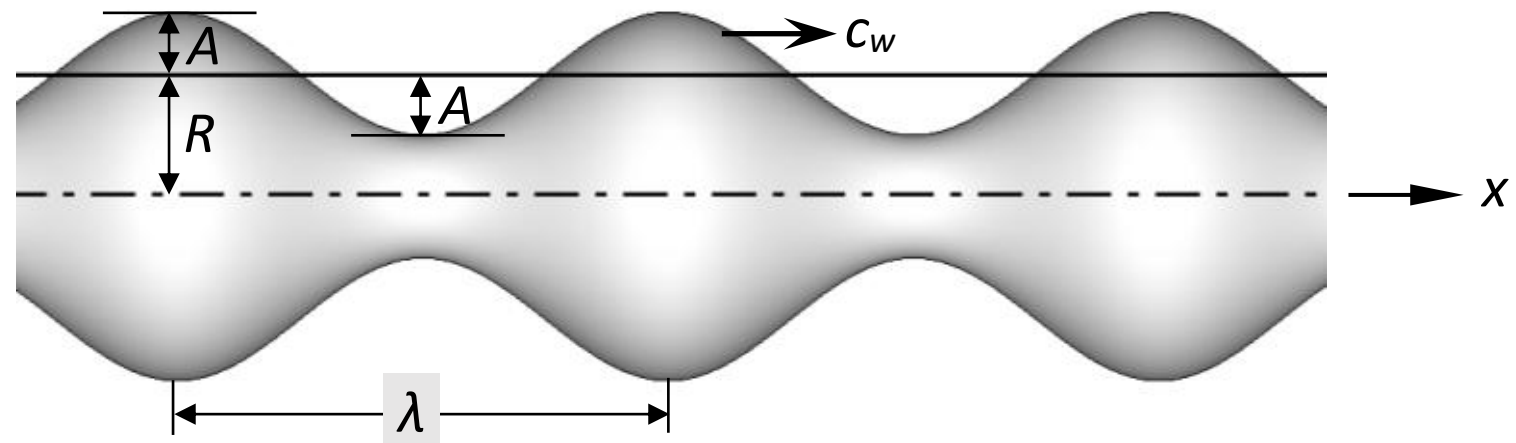

FIG. 1. Physical configuration of peristaltic-wave-driven flow in a circular tube. The tube wall is represented by an infinite sinusoidal wave train characterized by a mean radius $R$, wave amplitude $A$, and wavelength $\lambda$. Periodic radial oscillation of the wall generates a travelling peristaltic wave propagating at speed $c_w$ in the positive axial direction. The resulting wall deformation drives fluid motion within the tube.

To facilitate the analysis of flow structures, we introduce a reference frame moving with the travelling wave. In this wave-fixed frame, the wall geometry remains stationary, while the wall acquires an additional uniform velocity $c_w$ in the negative axial direction. Consequently, the velocity of every point on the wall is purely tangential to the wall surface and no normal wall motion exists.

To characterize the effect of fluid inertia, we define the Reynolds number as

$$Re = \frac{c_w R}{\nu} \tag{2.3}$$

where $\nu$ is the kinematic viscosity of the fluid. This definition differs from those commonly adopted in previous peristaltic-flow studies [10,12,14,15], where the wavelength is often included in the Reynolds number. The present definition is chosen to facilitate direct comparisons among flows with different wavelengths while maintaining a consistent measure of inertial effects.

To generalize the results, all analyses are performed in dimensionless form. The tube radius $R$ and wave speed $c_w$ are used as the characteristic length and velocity scales, respectively. Dimensionless variables are denoted by a tilde,

$$\tilde{\mathbf{x}} = \frac{\mathbf{x}}{R}, \quad \tilde{\mathbf{u}} = \frac{\mathbf{u}}{c_w} \tag{2.4}$$

Two key geometric parameters are introduced,

$$\chi = \frac{A}{R}, \quad \Lambda = \frac{\lambda}{R} \tag{2.5}$$

which represent the dimensionless wave amplitude and wavelength, respectively. The dimensionless problem is therefore completely characterized by three governing parameters: Re, $\chi$, and $\Lambda$.

The present study focuses on flows spanning the Stokes limit ($Re = 0$), low-inertia regimes ($Re = O(10)$), and moderate-inertia regimes ($Re = O(100)$). Within this parameter range, the flow remains predominantly axisymmetric and can be accurately described by axisymmetric dynamics. We note that at sufficiently short wavelengths and higher Reynolds numbers, symmetry breaking may occur, leading to non-axisymmetric flow structures and additional physical mechanisms. Such phenomena lie beyond the scope of the present work and will be addressed separately. The wave amplitude considered in this study covers the full range $0 < \chi < 1$, while the dimensionless wavelength spans $1 \leq \Lambda \leq 6$. Further increases in wavelength do not introduce qualitatively new flow behavior. The present parameter space is therefore sufficient to capture the essential flow regimes and transport mechanisms of interest.

Because the Reynolds numbers considered in the present study are relatively low and the travelling-wave geometry is spatially periodic, the flow is assumed to reach a steady state in the wave-fixed frame, with the same spatial periodicity as the wall deformation. This formulation has been widely employed in theoretical, numerical, and experimental studies of peristaltic transport [4,9,11,13-15]. Consequently, the computational domain consists of the entire three-dimensional fluid region contained within a single wavelength. Periodic boundary conditions are imposed on the two axial boundaries normal to the tube centerline, thereby representing an infinite train of travelling waves.

Within most of the parameter space considered in this study, the flow remains axisymmetric despite the fully three-dimensional computational domain. Only for the shortest wavelength considered ($\Lambda = 1$) at sufficiently large Reynolds numbers ($Re > 50$) does the flow lose axisymmetry and develop three-dimensional unsteady structures. Since the objective of the present work is to elucidate the mechanisms governing bolus formation in axisymmetric peristaltic-wave-driven flows, the analysis is restricted to the axisymmetric flow regime.

## III. NUMERICAL METHODS

In this study, we developed a numerical model framework based on the lattice Boltzmann method (LBM) to simulate the fluid dynamics driven by peristaltic waves. Compared with finite-difference and finite-volume methods, LBM is well suited for problems involving complex moving boundaries and large-scale parallel computation [27–30]. In LBM, the dependent variable is the particle distribution function $\mathbf{f}(\mathbf{x}, t)$, which quantifies the probability of finding an ensemble of molecules at position $\mathbf{x}$ with velocity $\mathbf{e}$ at time $t$. Continuum-level velocity $\mathbf{u}(\mathbf{x}, t)$ and density $\rho(\mathbf{x}, t)$ are obtained from moments of $\mathbf{f}(\mathbf{x}, t)$ over velocity space. In three dimensions, the velocity vector $\mathbf{e}$ can be discretized into 15, 19, or 27 components (referred to as D3Q15, D3Q19 and D3Q27) [27]. The D3Q19 model is adopted because it provides sufficient accuracy for the Reynolds numbers considered here while reducing computational cost.

The LBM equation with the Bhatnagar-Gross-Krook [30] representation for the collision operator is written as [28],

$$\mathbf{f}(\mathbf{x} + \mathbf{e}\delta_t, t + \delta_t) - \mathbf{f}(\mathbf{x}, t) = -\frac{1}{\tau}\left(\mathbf{f}(\mathbf{x}, t) - \mathbf{f}^{eq}(\mathbf{x}, t)\right) \tag{3.1}$$

where the discretized velocity vector $\mathbf{e}$ for D3Q19 is,

$$e_\alpha = \begin{cases} (0,0,0) & \alpha = 0 \\ (\pm 1,0,0), (0, \pm 1,0), (0,0, \pm 1) & \alpha = 1,2, \cdots ,6 \\ (\pm 1, \pm 1,0), (\pm 1,0, \pm 1), (0, \pm 1, \pm 1) & \alpha = 7,8, \cdots ,18 \end{cases} \tag{3.2}$$

where the subscript $\alpha$ is the index of components of velocity vector $\mathbf{e}$. The left-hand side (LHS) of equation (3.1) describes "streaming," the exchange of momentum between neighboring lattice nodes as a result of bulk advection and molecular diffusion. The right-hand side (RHS) describes the mixing, or collision of molecules that drive the distribution function ($\mathbf{f}(\mathbf{x},t)$) toward the equilibrium distribution function ($\mathbf{f}^{\mathrm{eq}}(\mathbf{x},t)$) with a relaxation time scale, τ. The relaxation time is linearly related to the fluid kinematic viscosity ν by

$$\nu = (2\tau - 1)c\delta x/6 \tag{3.3}$$

where $\delta x$ is the lattice spacing and $c = \delta x/\delta t$ is the basic speed on the lattice. In the low Mach number limit, the equilibrium distribution for the D3Q19 model is [27]

$$\mathbf{f}^{eq}(\mathbf{x},t) = w_\alpha \rho(\mathbf{x},t)\left[1 + 3\frac{\mathbf{e}_\alpha\cdot\mathbf{u}}{c^2} + \frac{9}{2}\frac{(\mathbf{e}_\alpha\cdot\mathbf{u})^2}{c^4} - \frac{3}{2}\frac{(\mathbf{u}\cdot\mathbf{u})}{c^2}\right] \tag{3.4}$$

where $w_\alpha$ are weighting factors, $w_0 = 1/3$, $w_\alpha = 1/18$ for $\alpha = 1 - 6$ and $w_\alpha = 1/36$ for $\alpha = 7 - 18$. The continuum-level fluid density $\rho(\mathbf{x},t)$ and momentum $\rho\mathbf{u}(\mathbf{x},t)$ are obtained from the discretized moments of the particle distribution function,

$$\rho(\mathbf{x},t) = \textstyle\sum_\alpha f_\alpha(\mathbf{x},t), \quad \rho(\mathbf{x},t)\mathbf{u}(\mathbf{x},t) = \sum_\alpha f_\alpha(\mathbf{x},t)\mathbf{e}_\alpha \tag{3.5}$$

In the treatment of moving walls, we use 2nd order accurate scheme advanced by Lallemand and Luo [31], Ladd [32], and Bouzidi et al. [33].

To obtain the Stokes-limit solutions ($Re = 0$), a linearized equilibrium distribution function was employed following Chun and Ladd [34]. By retaining only the first-order velocity terms in the equilibrium distribution, the recovered macroscopic equations reduce to the Stokes equations, and fluid inertia is eliminated identically.

The numerical model has been extensively validated in our previous studies over a wide range of parameters [35-37]. For the present problem, additional comparisons with published results will be presented in the following sections.

## IV. RESULT VALIDATION

A grid-independence study was performed. Increasing the Reynolds number and decreasing the wavelength both lead to increasingly localized flow features, thereby imposing more stringent requirements on spatial resolution. The case with $Re = 100$ and $\Lambda = 2$, was selected as the most demanding test case for grid convergence.

Three grid resolutions were examined, with $n_r = 60$, 80, and 100 grid points spanning the mean tube radius $R$, corresponding to coarse, medium, and fine resolutions, respectively. To assess grid convergence, the axial velocity $\tilde{u}_x$ in the frame of reference fixed to the wave was compared on two representative cross-sections located at the wave crest and wave trough. For reference, Fig. 2(a) presents contours of $\tilde{u}_x$ on a longitudinal section obtained using the medium-resolution grid ($n_r = 80$), providing an overview of the flow field. Figure 2(b) compares the radial profiles of $\tilde{u}_x$ at the wave-crest and wave-trough cross-sections for the three grid resolutions.

The results obtained with different resolutions are in close agreement. The largest discrepancy among the three solutions occurs at the centerline of the wave-trough cross-section, where the channel reaches its minimum width. At this location, the deviation of the coarse-grid solution from the fine-grid solution is approximately 2%, whereas the deviation of the medium-grid solution remains below 0.5%. Considering both computational accuracy and efficiency, the medium-resolution grid ($n_r = 80$) was adopted for all simulations at $Re = 100$.

Similar grid-convergence tests were also conducted for the remaining Reynolds numbers considered in this study. In all cases, the selected grid resolution yielded deviations below 1% relative to the corresponding fine-grid solutions. The grid resolutions employed for each Reynolds number are summarized in Table 1.

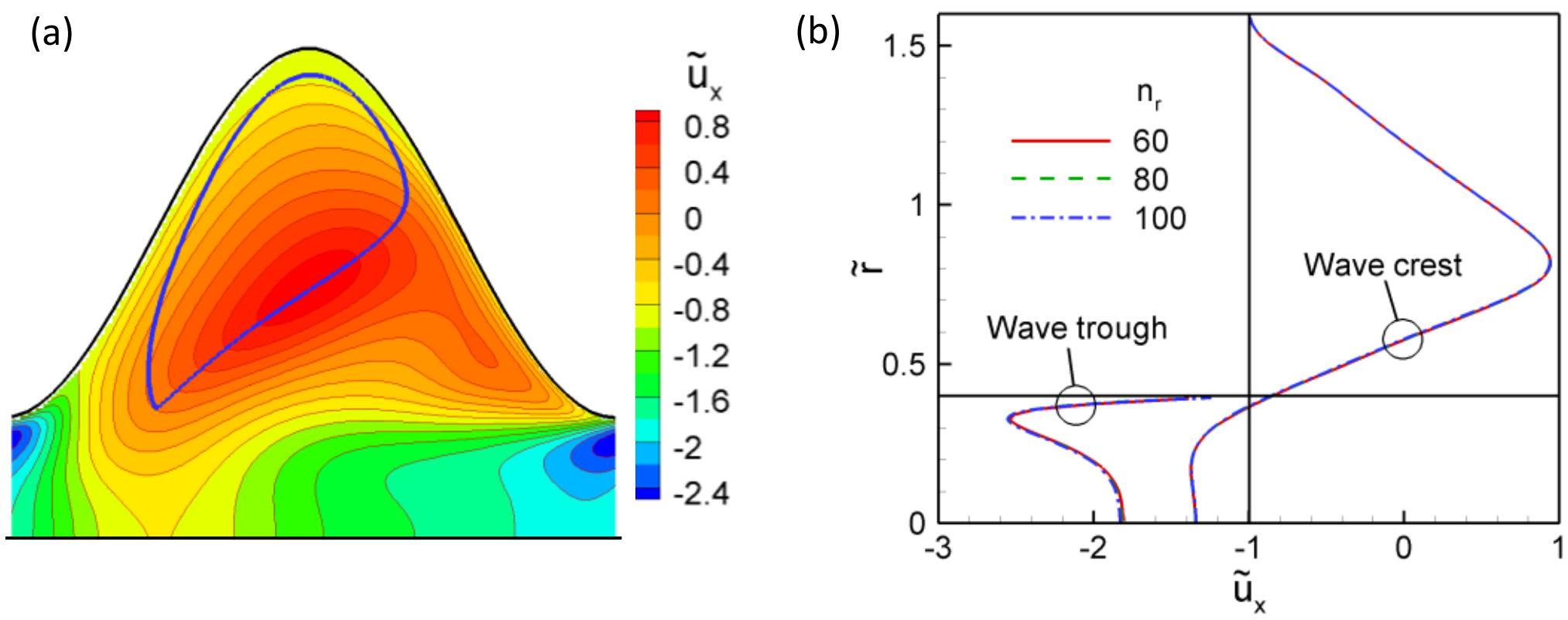

FIG. 2. Effect of grid resolution on axial velocity ($\tilde{u}_x$) with $\Lambda = 2$ and $\chi = 0.6$ for $Re = 100$. (a) Iso-contours of $\tilde{u}_x$ on a longitudinal section for $n_r = 80$. The closed blue line indicates the boundary of bolus. (b) Radial profiles of $\tilde{u}_x$ at the wave crest and wave trough for $n_r = 60$, 80, and 100.

Table 1. Number of grid points used to resolve the mean tube radius $R$ for different Reynolds numbers.

| $Re$ | 0 | 10 | 20 | 40 | 50 | 70 | 100 |
|---|---|---|---|---|---|---|---|
| $n_r$ | 30 | 40 | 40 | 50 | 50 | 70 | 80 |

## V. RESULTS AND DISCUSSION

For all Reynolds numbers and wave parameters considered in this study, the flow reaches a steady state in the wave-fixed frame, and all analyses presented below are performed under this condition.

### A. Fluid trapping and wavelength-normalized flow similarity in the Stokes limit

We first examine the Stokes flow in the zero-Reynolds-number limit. In the reference frame moving with the travelling wave, the tangential motion of the wall drives a globally connected through-flow predominantly in the negative $x$-direction. At relatively small wave amplitudes ($\chi$), the streamlines follow the curved wall smoothly and remain approximately symmetric about the wave crest. As $\chi$ increases, the progressive narrowing near the wave trough produces increasingly strong geometric confinement. The resulting confinement progressively enhances the viscous resistance of the constricted passage and drives a pressure rise inside the enlarged wave cavity beneath the wave crest. Once the resulting adverse pressure gradient becomes sufficiently strong, a centerline return-flow region emerges within the surrounding negative-$x$ through-flow, causing the globally connected streamline topology to break down and a separatrix-enclosed bolus to form. Such symmetric trapping structures have been extensively reported in previous studies of inertia-free peristaltic transport [4,9,13,14]. Because the travelling-wave geometry is symmetric about the wave crest and the Stokes momentum equations preserve spatial reversibility in the absence of inertia, the resulting flow structure remains approximately symmetric with respect to the wave crest. As $\chi$ increases further, the pressure-driven return flow strengthens continuously, causing the separatrix to expand and the bolus size to increase correspondingly, consistent with classical observations of peristaltic trapping [4,11,12,14].

Figure 3 illustrates a representative Stokes-flow structure for $\Lambda = 4$ and $\chi = 0.4$. The streamline topology shown in Fig. 3(a) reveals the symmetric trapped bolus beneath the wave crest. The axial velocity contours in Fig. 3(b) show strong negative velocities near the wave trough together with positive velocities near the centerline inside the wave cavity, directly revealing the centerline return-flow structure associated with trapping. Meanwhile, the radial velocity contours in Fig. 3(c) reveal the radial transport responsible for closing the recirculating streamline structure.

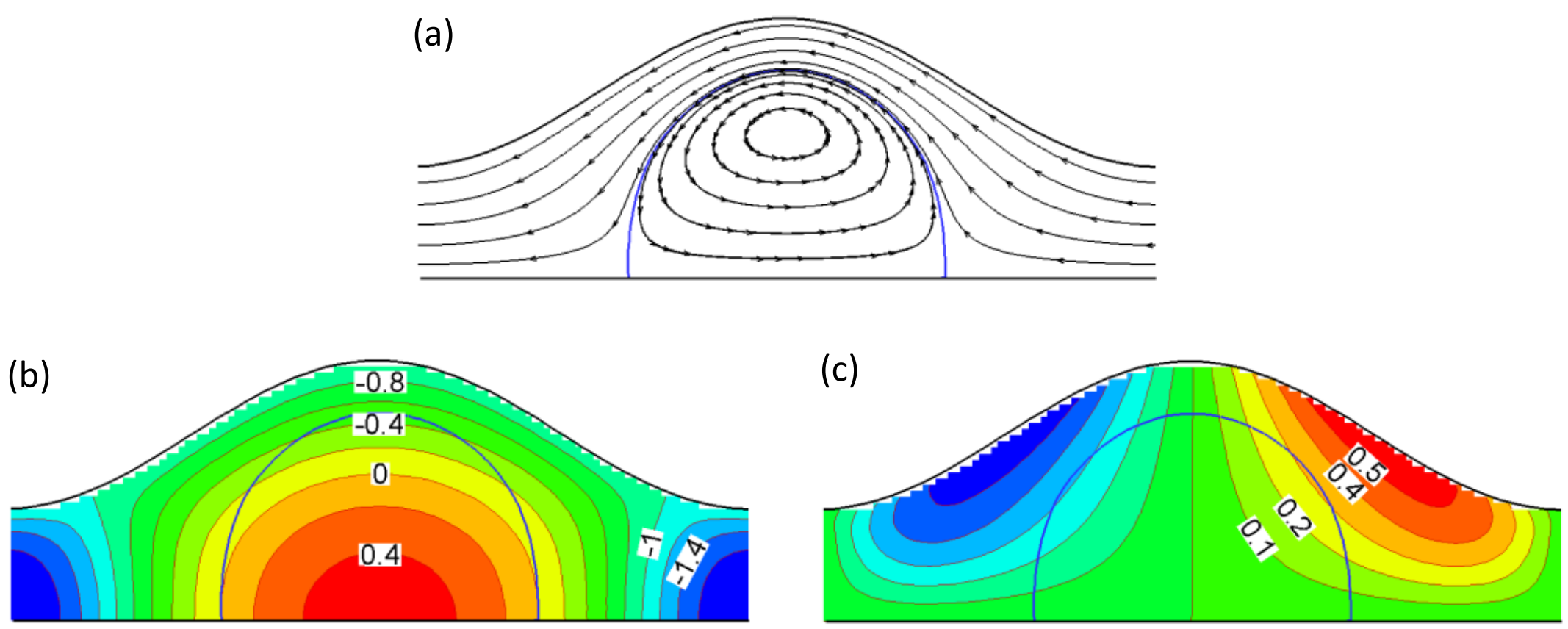

FIG. 3. Flow structure in the Stokes limit for $\Lambda = 4$ and $\chi = 0.4$. (a) Streamlines in the wave-fixed reference frame, showing the formation of a trapped bolus beneath the wave crest. The blue curve denotes the trapping boundary separating the recirculating bolus region from the surrounding through-flow. (b) Iso-contours of the dimensionless axial velocity component $\tilde{u}_x$. Positive axial velocities dominate the bolus core, while reverse flow develops near the wave trough. (c) Iso-contours of the dimensionless radial velocity component $\tilde{u}_r$, illustrating the upward and downward motions responsible for the recirculating flow organization.

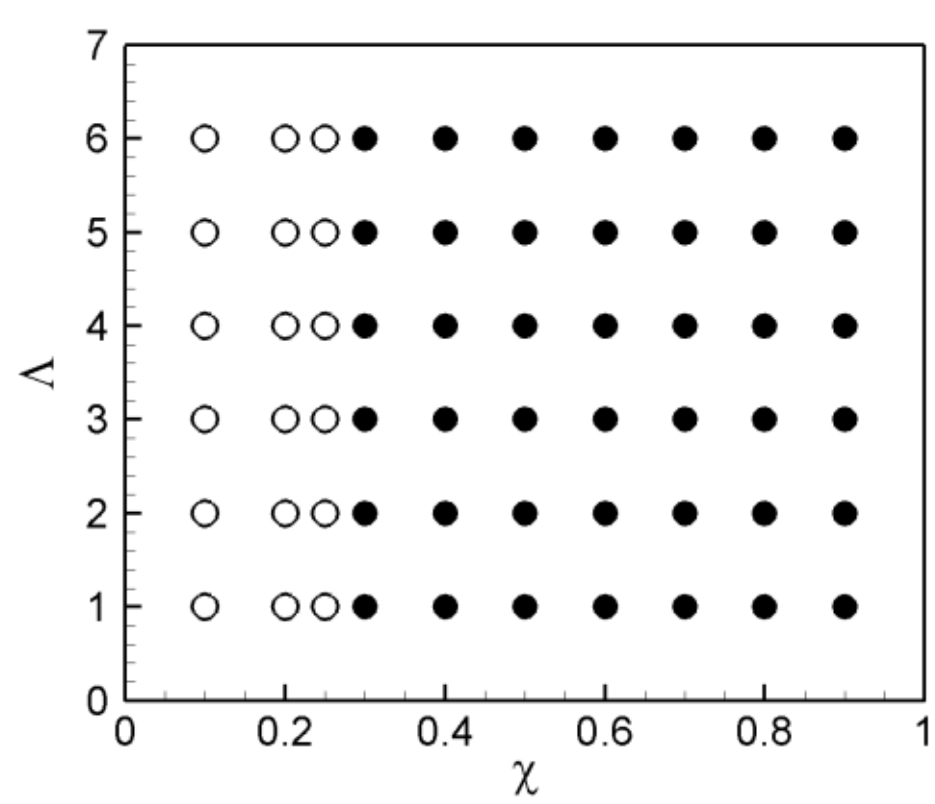

FIG. 4. Bolus formation regime diagram in the Stokes limit. Distribution of trapped and non-trapped flow states in the two-dimensional parameter space spanned by the dimensionless wavelength $\Lambda$ and wave amplitude $\chi$. Filled circles denote cases in which a trapped bolus is present, while open circles indicate flows without bolus formation.

Figure 4 presents the bolus-formation regime diagram over the parameter ranges $1 \leq \Lambda \leq 6$ and $0.1 \leq \chi \leq 0.9$. Despite the substantial variation in wavelength, bolus formation consistently occurs within a relatively narrow range of wave amplitudes, approximately $0.25 < \chi < 0.3$. This weak dependence of the trapping threshold on $\Lambda$ indicates that, in the Stokes limit, bolus formation is governed primarily by the wave amplitude, whereas the influence of wavelength on trapping onset remains comparatively weak. More importantly, the nearly wavelength-insensitive trapping threshold suggests that the underlying flow organization retains a substantial degree of similarity across different wavelengths, despite the large variation in the geometric scale of the travelling wave.

As an important quantitative characterization of the flow organization, Fig. 5 presents the variation of the dimensionless bolus volume $\tilde{V}_{bolus}$ $(= V_{bolus}/R^3)$ with wave amplitude $\chi$ for different wavelengths $\Lambda$, together with the corresponding bolus volume fraction $\beta$,

$$\beta = V_{bolus}/V_{wave} = V_{bolus}/\int_0^{\lambda}\int_0^{h(x)} 2\pi r dr dx \tag{5.1}$$

As shown in Fig. 5(a), for all wavelengths considered here, the bolus volume emerges only after $\chi$ exceeds a nearly common critical value, consistent with the trapping regime diagram shown previously in Fig. 4. Beyond this threshold, the bolus volume increases continuously with increasing wave amplitude. As $\chi$ approaches unity, the throat region near the wave trough approaches complete occlusion and the bolus correspondingly expands to occupy nearly the entire flow region enclosed within one wavelength. Despite the substantial variation in wavelength, the evolution curves corresponding to different $\Lambda$ retain a remarkable degree of similarity.

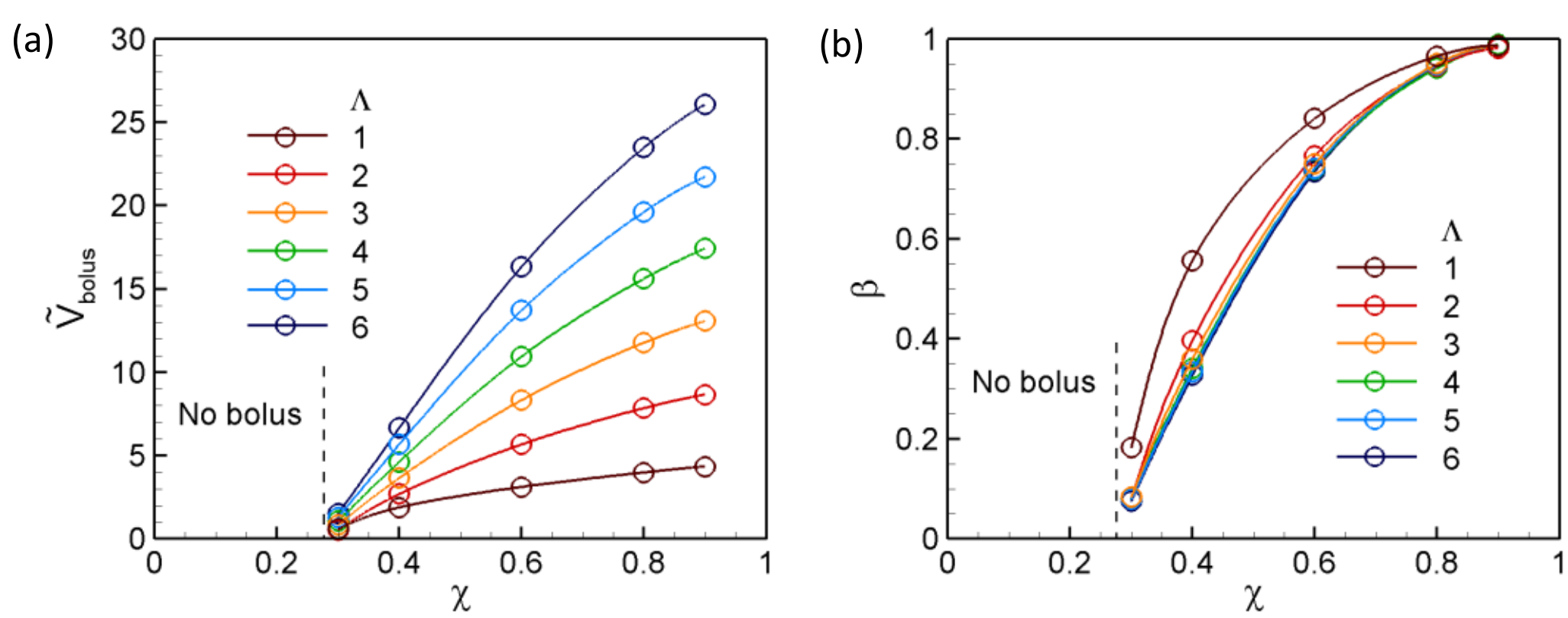

FIG. 5. Bolus volume and bolus volume fraction in the Stokes limit.
(a) Dimensionless bolus volume $\tilde{V}_{bolus}$ as a function of wave amplitude $\chi$ for different wavelengths $\Lambda$. Bolus formation occurs only above a critical wave amplitude, consistent with the regime diagram shown previously. Increasing wavelength substantially enlarges the bolus volume and promotes the growth of trapped regions. (b) Ratio between the bolus volume and the total fluid volume, $\beta$, as a function of wave amplitude $\chi$. As the wavelength increases, the curves progressively collapse toward a common asymptotic behavior corresponding to the infinite-wavelength limit.

This similarity becomes even more evident in Fig. 5(b), which shows the variation of the bolus volume fraction $\beta$ with $\chi$. As $\Lambda$ increases, the corresponding curves progressively converge toward a common asymptotic behavior, which evidently corresponds to the infinite-wavelength limit. Except for the shortest-wavelength case ($\Lambda = 1$), the remaining curves remain closely clustered over a broad range of wave amplitudes, indicating that the overall trapping structure retains a substantial degree of similarity across different wavelengths in the Stokes limit. At the same time, the noticeably larger bolus volume fraction observed for ($\Lambda = 1$) suggests that wavelength effects become increasingly important in the short-wavelength regime, where the increasingly strong geometric bending of the wave progressively modifies the trapping structure and promotes the growth of the trapped region.

Fluid transport represents another important aspect of peristaltic-wave-driven flows. From the laboratory reference frame, the travelling wave carries fluid downstream and produces a net positive-$x$ transport. In the reference frame moving with the travelling wave, however, the tangential wall motion generates a relative backward fluid transport in the negative $x$-direction. We therefore characterize the transport behavior using two dimensionless volumetric transport rates: the laboratory-frame carrying transport rate, $\tilde{Q}_{carry}$, and the wave-frame slip transport rate, $\tilde{Q}_{slip}$.

$$\tilde{Q}_{carry} = \frac{1}{T}\int_0^T \int_0^{h(x',t)} u_x'(x',r,t) 2\pi r dr\, dt / cR^2 \tag{5.2}$$

where $u_x'$ is the axial velocity in the laboratory reference frame, $x'$ denotes an arbitrary fixed axial position in the laboratory frame, and $T = \lambda/c$ is the wave period. This quantity characterizes the net downstream fluid transport induced by the travelling wave.

$$\tilde{Q}_{slip} = -\int_0^{h(x)} u_x(x,r) 2\pi r dr\, / cR^2 \tag{5.3}$$

where $u_x(x,r)$ is the axial velocity in the wave-fixed reference frame. Physically, $\tilde{Q}_{slip}$ characterizes the relative backward slip transport of the fluid with respect to the travelling wave.

$$\tilde{Q}_{carry} + \tilde{Q}_{slip} = \frac{1}{T}\int_0^{\lambda} \int_0^{h(x)} 2\pi r dr\, dx / R^3 \tag{5.4}$$

indicating that the sum of the carrying and slip transport rates is determined solely by the fluid volume enclosed within a single wavelength and the wave period. Physically, the right-hand side represents the dimensionless volumetric flow rate associated with transporting the entire fluid volume contained within one wavelength over a single wave period.

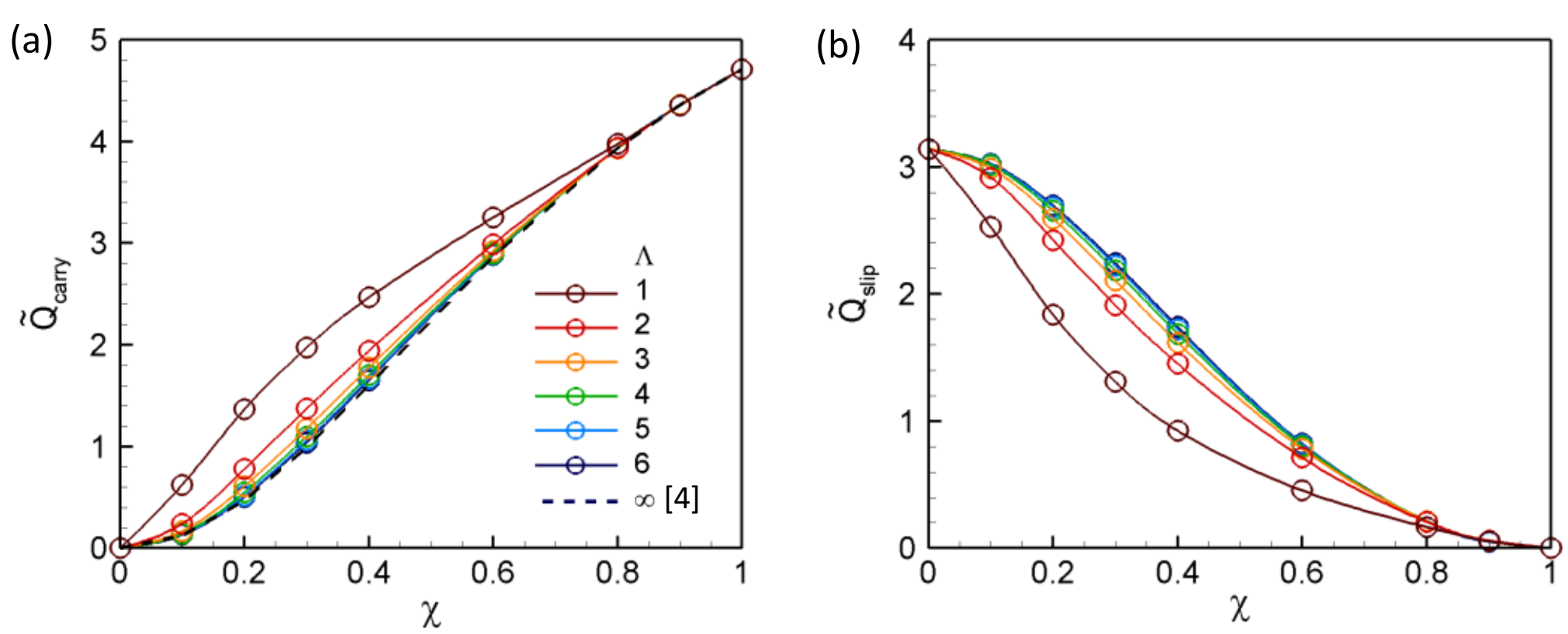

FIG. 6. Variation of the dimensionless volumetric flow rates with wave amplitude $\chi$ for different wavelength ratios $\Lambda$ in the Stokes limit. (a) Dimensionless carrying transport rate $\tilde{Q}_{carry}$ in the laboratory reference frame, representing the net downstream fluid transport induced by the travelling wave. (b) Dimensionless slip transport rate $\tilde{Q}_{slip}$ in the wave-fixed reference frame, representing the relative upstream fluid transport with respect to the travelling wave. Symbols denote the present numerical results, while the dashed curve represents the long-wavelength analytical prediction of Shapiro et al. [4].

Figure 6 presents the variation of the two transport rates with wave amplitude $\chi$ for different wavelengths $\Lambda$. For all wavelengths considered here, $\tilde{Q}_{carry}$ increases monotonically with increasing $\chi$, indicating progressively stronger downstream fluid transport induced by the travelling wave. In contrast, $\tilde{Q}_{slip}$ decreases continuously with increasing $\chi$, reflecting the progressive suppression of the backward slip transport in the wave-fixed reference frame as the trapped region expands. For comparison, the long-wavelength analytical solution of Shapiro et al. [4] is also included in Fig. 6. The excellent agreement observed for the largest wavelength considered provides an independent validation of the present numerical methodology and confirms the convergence toward the classical long-wavelength asymptote. Both transport measures exhibit a substantial degree of similarity across different wavelengths. As $\Lambda$ increases, the evolution curves of both $\tilde{Q}_{carry}$ and $\tilde{Q}_{slip}$ progressively converge toward common asymptotic behaviors corresponding to the infinite-wavelength limit. This asymptotic convergence further confirms that, despite the large variation in wavelength, the underlying transport structure retains a substantial degree of similarity in the Stokes limit.

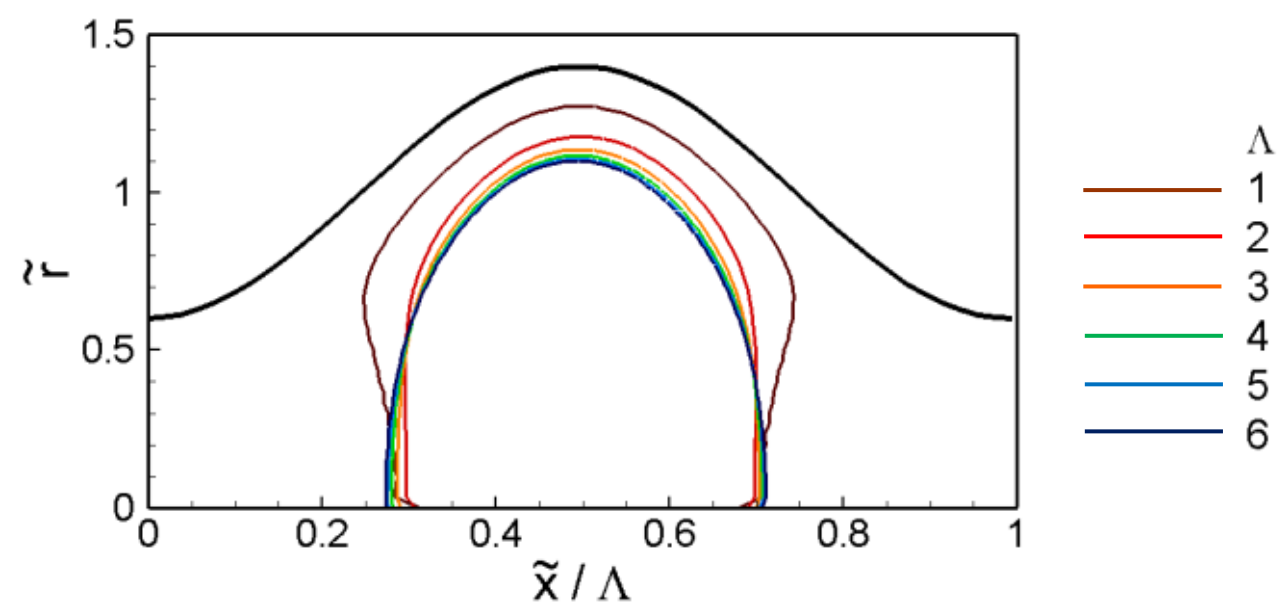

FIG. 7. Comparison of bolus shapes for different wavelengths $\Lambda$ in the Stokes limit at $\chi = 0.4$. The axial coordinate is normalized by the wavelength, $\tilde{x}/\Lambda$, to compare the wavelength-normalized bolus structures. As the wavelength increases, the bolus boundaries progressively converge toward a common asymptotic shape, indicating the existence of an underlying similarity structure in the axial flow organization of inertia-free peristaltic flows. The black curve denotes the tube wall geometry.

Figure 7 further elucidates the origin of the similarity observed across different wavelengths. The figure presents the bolus boundaries for $1 \leq \Lambda \leq 6$ at a fixed wave amplitude $\chi = 0.4$, where the axial coordinate is normalized by the wavelength. As shown in Fig. 7, the normalized bolus shapes progressively collapse toward a common profile as $\Lambda$ increases. Similar trends are also observed in the corresponding velocity distributions, further confirming that the underlying flow structures become increasingly wavelength-independent in the long-wavelength Stokes limit.

This similarity can be understood directly from the nondimensional Stokes equations. When the axial coordinate is scaled by the wavelength and the radial coordinate is scaled by the mean tube radius,

$$\hat{x} = \frac{x}{\lambda}, \quad \hat{r} = \frac{r}{R} \tag{5.5}$$

the axial and radial derivatives scale as

$$\frac{\partial}{\partial x} \sim \frac{1}{\lambda}, \quad \frac{\partial}{\partial r} \sim \frac{1}{R} \tag{5.6}$$

and therefore the Laplacian operator becomes

$$\nabla^2 \sim \frac{1}{R^2}\left(\frac{\partial^2}{\partial \hat{r}^2} + \Lambda^{-2}\frac{\partial^2}{\partial \hat{x}^2}\right) \tag{5.7}$$

The essential feature revealed by this scaling is that the axial viscous coupling enters the Stokes equations only through the small parameter $\Lambda^{-2}$. As $\Lambda$ increases, the influence of axial viscous diffusion progressively weakens and the flow approaches a quasi-local long-wavelength regime in which the leading-order flow organization is governed primarily by the local wall geometry. The explicit wavelength dependence therefore appears only as a higher-order correction of $O(\Lambda^{-2})$. This scaling directly explains the progressive collapse of the normalized bolus boundaries and velocity fields observed in Fig. 7. Once the wavelength becomes sufficiently long, the flow adjusts locally to the slowly varying wall geometry, causing the separatrix location and the overall recirculating structure to become increasingly insensitive to the global wavelength. In contrast, at smaller $\Lambda$, the stronger axial variation of the wall geometry enhances axial viscous coupling and produces increasingly visible deviations from the long-wavelength similarity regime.

### B. Breakdown of wavelength-normalized similarity at weak inertia

In the presence of fluid inertia, the fluid tends to maintain its motion rather than moving entirely under viscous control along the solid boundary. As a result, in the wave-fixed reference frame, fluid passing leftward through the narrow throat near the wave trough continues to move leftward along the axis after leaving the throat. This motion impinges on the trapped bolus, causing the trapping structure to shrink and, under certain conditions, disappear altogether. On the other hand, increasing inertia also promotes effects that are suppressed by viscous forces in the Stokes limit, particularly those associated with the strongly curved geometry of short-wavelength waves. We first consider the weak-inertia case, $Re = 20$.

Figure 8 presents the bolus-formation regime diagram for the weak-inertia case, $Re = 20$. Compared with the Stokes-flow results shown in Fig. 4, the presence of inertia suppresses bolus formation in the long- and intermediate-wavelength regimes. As the wavelength decreases from $\Lambda = 6$ to $\Lambda = 2$, the critical wave amplitude required for bolus formation increases monotonically from approximately $\chi = 0.25$ to nearly $\chi = 0.4$. Specifically, bolus formation remains confined to $0.25 < \chi < 0.3$ for $\Lambda = 6$ and 5, shifts to $0.3 < \chi < 0.35$ for $\Lambda = 4$ and 3, and further increases to $0.35 < \chi < 0.4$ for $\Lambda = 2$. These results suggest that the classical trapping structure becomes increasingly susceptible to disruption as the wavelength decreases.

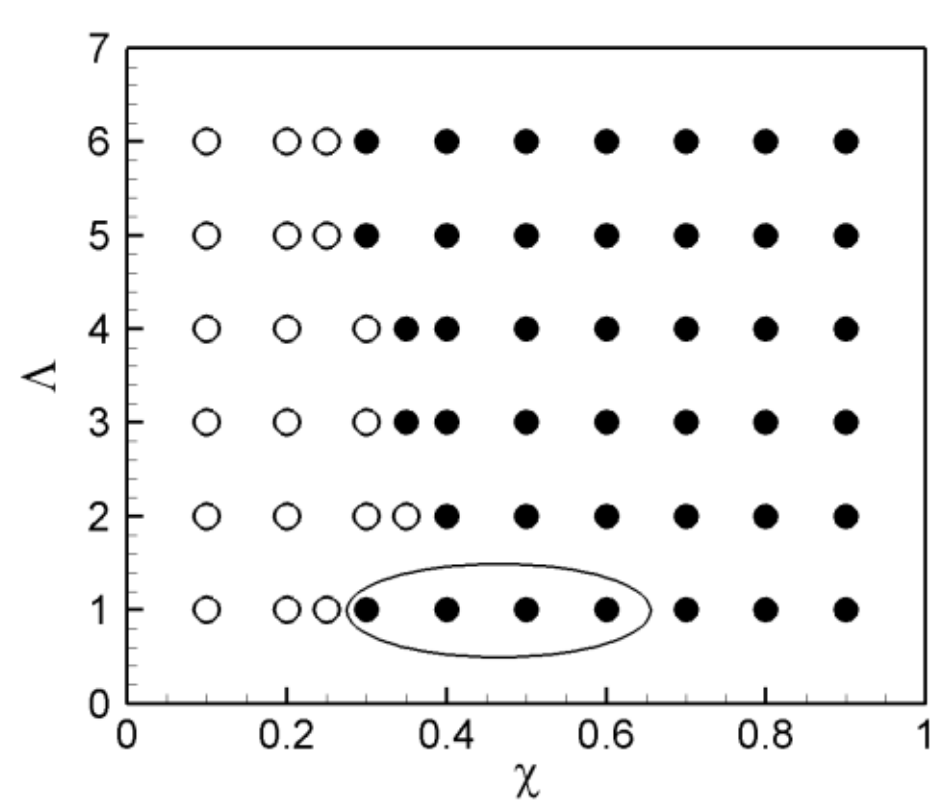

FIG. 8. Regime diagram of bolus formation for the weak-inertia case ($Re = 20$) over the parameter ranges $1 \leq \Lambda \leq 6$ and $0.1 < \chi < 0.9$. Filled symbols denote cases in which bolus formation occurs, while open symbols represent flows without trapping. The circled region approximately identifies the short-wavelength regime in which the strong geometric bending of the wave becomes sufficient to induce bolus formation.

Since the bolus length scales approximately with wavelength, shorter wavelengths produce progressively smaller recirculating regions, which are more readily penetrated by the through-flow emerging from the throat. As a result, the classical bolus shrinks and eventually disappears at intermediate wavelengths.

A strikingly different behavior emerges when the wavelength is further reduced to $\Lambda = 1$. In this case, bolus formation again occurs within the range $0.25 < \chi < 0.3$, apparently suggesting that inertia no longer suppresses trapping. However, the flow structure responsible for trapping differs substantially from that observed at longer wavelengths. At such short wavelengths, the strongly curved wave geometry produces pronounced turning of fluid trajectories near the wavy boundary. In the Stokes limit, this effect is largely masked by viscous diffusion. With the introduction of inertia, however, the curvature-induced flow structure becomes increasingly pronounced, leading to the formation of an off-axis bolus within the highly curved region of the wave. In three dimensions, this structure takes the form of a toroidal, donut-like bolus.

To further elucidate the underlying trapping mechanisms, Fig. 9 presents the streamline patterns at $Re = 20$ for wave amplitude ratios $\chi = 0.3$ (left column) and $\chi = 0.4$ (right column) over a range of wavelengths. The case $\chi = 0.3$ is particularly informative because it lies close to the trapping threshold for all wavelengths considered. Under these conditions, some wavelengths exhibit bolus formation while others do not, allowing the differences between the underlying trapping mechanisms to be clearly distinguished.

For $\chi = 0.3$, a classical axis-centered bolus is observed at the longest wavelengths $\Lambda = 6$ and 5. As the wavelength decreases, the influence of the throat flow on the bolus becomes increasingly pronounced, causing the bolus to shrink progressively. For $\Lambda = 4$, 3, and 2, the bolus disappears completely. Surprisingly, when the wavelength is further reduced to $\Lambda = 1$, a new bolus reappears. Unlike the classical trapping structure, this bolus is detached from the centerline and resides within the highly curved region of the travelling wave. Similar nonaxial trapping states have previously been reported in two-dimensional peristaltic channels at sufficiently large wave numbers [38,39], suggesting that the short-wavelength trapping observed here may belong to a distinct trapping regime.

The distinction between these two trapping modes becomes even clearer when compared with the higher-amplitude case, $\chi = 0.4$ (right column). Unlike the $\chi = 0.3$ case, where the two trapping modes are clearly separated, boluses are present for all wavelengths considered. As the wavelength decreases from $\Lambda = 6$ to 3, the bolus gradually shrinks under the influence of the throat flow, consistent with the behavior observed for $\chi = 0.3$. However, when the wavelength is further reduced to $\Lambda = 2$ and 1, the bolus moves away from the centerline and becomes increasingly associated with the highly curved region of the wave. More importantly, the bolus at $\Lambda = 1$ is larger than that at $\Lambda = 2$. If the trapping behavior were governed solely by the classical long-wavelength mechanism, a further reduction in wavelength would be expected to continue suppressing the bolus. The observed increase in bolus size therefore indicates that a different trapping mechanism becomes increasingly important in the shortest-wavelength regime.

Figure 10 presents the dimensionless bolus volume $\tilde{V}_{bolus}$ and the corresponding bolus volume fraction $\beta$ as functions of wave amplitude $\chi$ for different wavelength ratios $\Lambda$. A clear distinction can be observed between the shortest-wavelength case, $\Lambda = 1$, and the longer-wavelength cases. The evolution curves for $\Lambda = 1$ differ qualitatively from those for $2 \le \Lambda \le 6$, further supporting the conclusion from Fig. 9 that trapping in the short-wavelength regime is governed by a different mechanism.

For the longer-wavelength cases $2 \le \Lambda \le 6$, the $\beta$ curves exhibit a substantial degree of similarity. Although noticeable differences exist at small $\chi$, the curves progressively converge as $\chi$ increases. This behavior indicates that the wavelength-normalized similarity identified in the Stokes limit is weakened, but not destroyed, by inertia. The convergence of the long-wavelength curves at larger amplitudes suggests that the classical trapping mechanism remains the dominant pathway governing bolus formation in this regime.

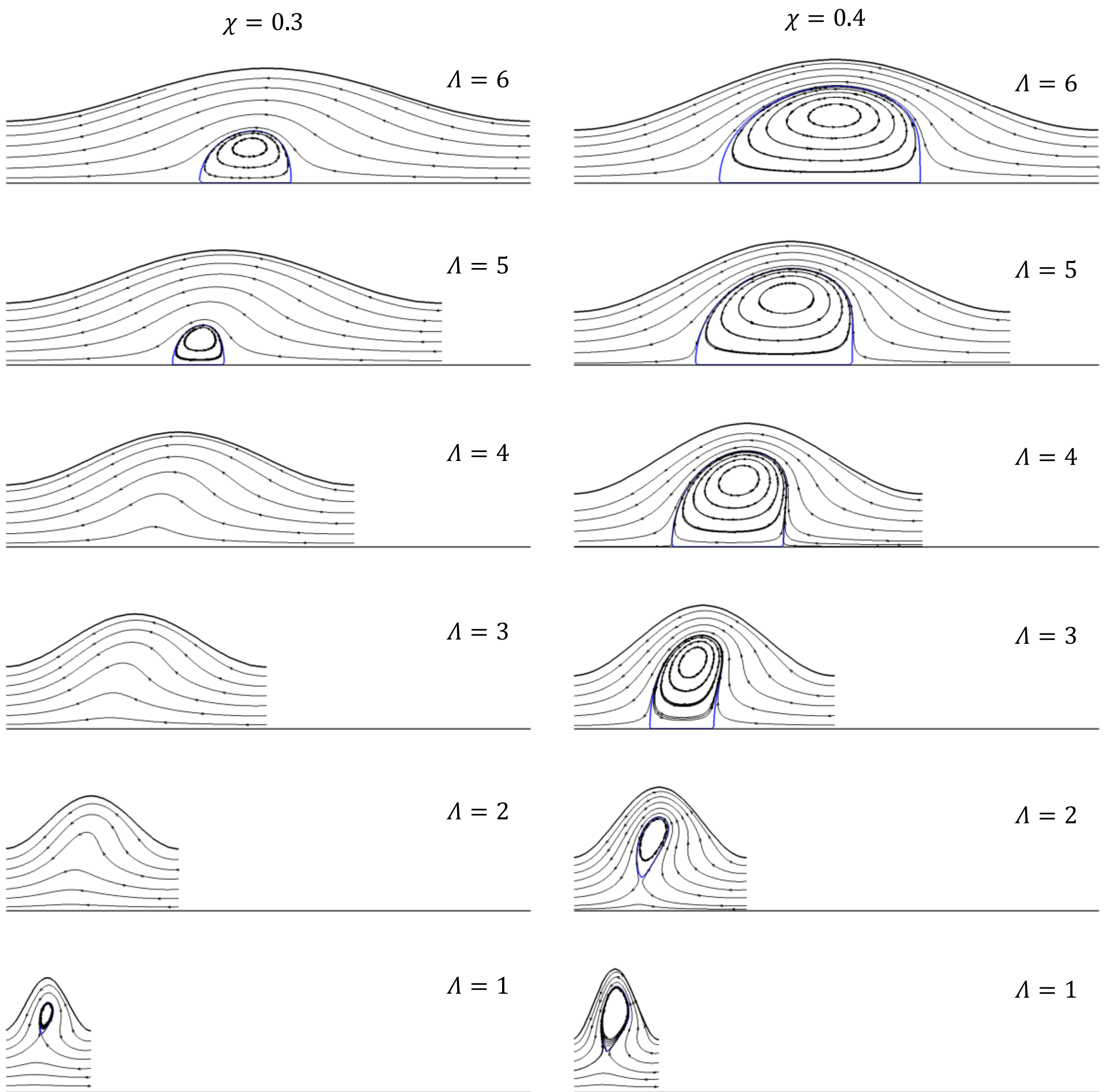

FIG. 9. Streamline patterns in the wave frame at $Re = 20$ for amplitude ratios $\chi = 0.3$ (left column) and $\chi = 0.4$ (right column). The wavelength ratio $\Lambda$ decreases from 6 to 1 from top to bottom. Blue contours indicate the bolus boundaries. For $\chi = 0.3$, the bolus undergoes substantial changes in size and location as the wavelength decreases, including its disappearance at intermediate wavelengths and reappearance at $\Lambda = 1$. For $\chi = 0.4$, boluses are present throughout the wavelength range considered, while their morphology and spatial distribution vary significantly with wavelength.

Figure 11 compares the dimensionless volumetric flow rates $\tilde{Q}_{carry}$ at $Re = 20$ with those obtained in the Stokes limit. In contrast to the substantial changes observed in bolus formation and trapping topology, the overall transport rate remains relatively insensitive to inertia for most wavelengths. For $2 \leq \Lambda \leq 6$, the $Re = 20$ results remain close to their Stokes-flow counterparts over the entire range of wave amplitudes considered. In contrast, the shortest-wavelength case $\Lambda = 1$ exhibits a noticeable deviation from the Stokes-flow behavior, particularly at small and moderate wave amplitudes. This deviation coincides with the emergence of the curvature-induced trapping mechanism identified in Figs. 9 and 10, suggesting that the curvature-induced trapping mechanism influences not only the trapping topology but also the overall transport characteristics in the short-wavelength regime.

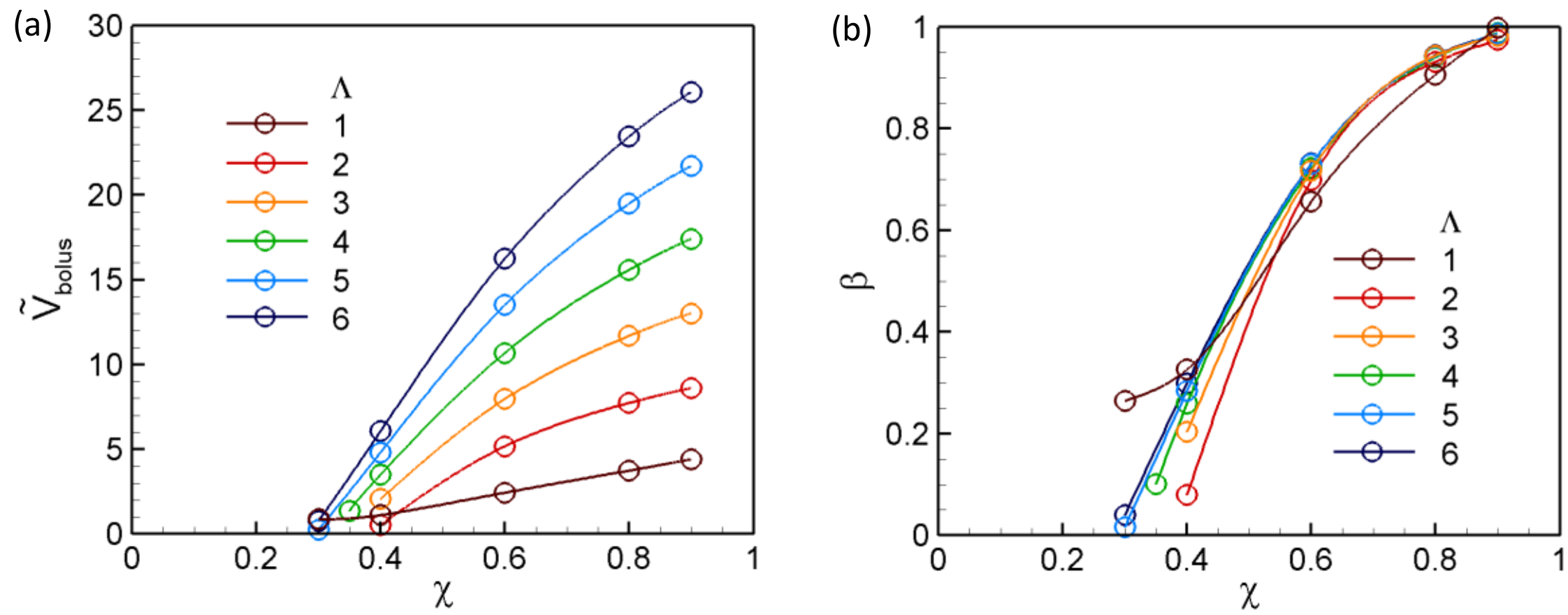

FIG. 10. Variation of the dimensionless bolus volume $\tilde{V}_{bolus}$ and the bolus volume fraction $\beta$ with wave amplitude $\chi$ at $Re = 20$ for different wavelengths $\Lambda$. (a) Dimensionless bolus volume $\tilde{V}_{bolus}$ as a function of $\chi$. (b) Bolus volume fraction $\beta$ as a function of $\chi$. For $\Lambda \geq 2$, the evolution curves exhibit substantial similarity and progressively approach a common trend with increasing wavelength, indicating that the long-wavelength similarity observed in the Stokes limit remains partially preserved under weak inertia. In contrast, the $\Lambda = 1$ case exhibits a qualitatively different evolution behavior, suggesting that bolus formation in the short-wavelength regime is governed by a different mechanism associated with strong geometric-bending effects.

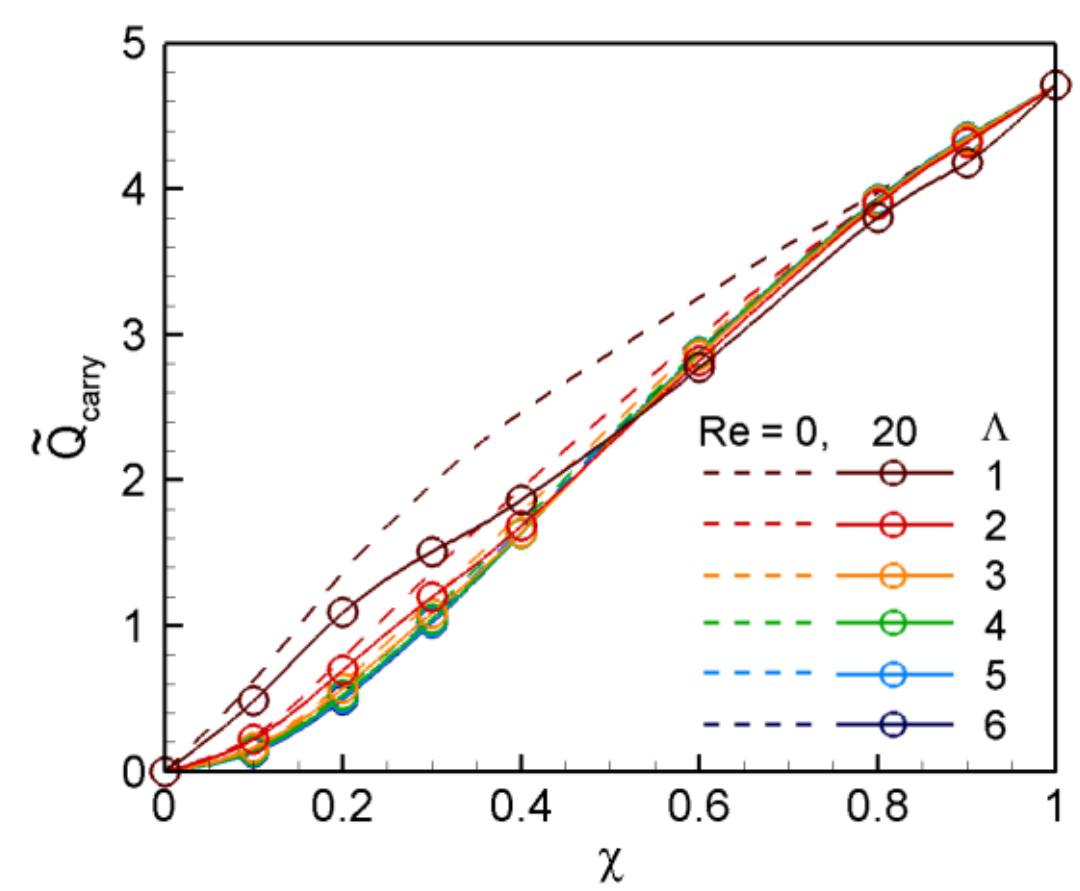

FIG. 11. Variation of the dimensionless volumetric flow rate $\tilde{Q}_{carry}$ with wave amplitude $\chi$ for different wavelength ratios $\Lambda$ at $Re = 20$. The corresponding Stokes-flow results ($Re = 0$, dashed lines) are included for comparison.

Taken together, Figs. 8–11 indicate that two distinct trapping mechanisms coexist at finite inertia. For the longer-wavelength cases $2 \leq \Lambda \leq 6$, bolus formation remains primarily governed by the classical trapping mechanism, and both the trapping characteristics and transport rates remain close to their Stokes-flow counterparts. In contrast, the qualitatively different behavior observed for $\Lambda = 1$, including the altered trapping topology, bolus-volume evolution, and transport characteristics, reflects the increasing importance of the curvature-induced trapping mechanism. These results suggest that the curvature-induced trapping mechanism becomes increasingly important in short-wavelength peristaltic transport once inertial effects are present.

### C. Interaction and merging of trapping pathways at moderate inertia

With increasing fluid inertia, both trapping mechanisms identified in the weak-inertia regime become more pronounced. On one hand, the stronger interaction between the throat flow and the trapped bolus further suppresses the classical trapping structure. On the other hand, inertia promotes the development of the curvature-induced trapping mechanism. As a result, the competition between the two mechanisms becomes increasingly important, and the influence of the curvature-induced trapping mechanism extends over a broader range of wavelengths and wave amplitudes. To investigate these effects, we next consider the moderate-inertia case, $Re = 100$.

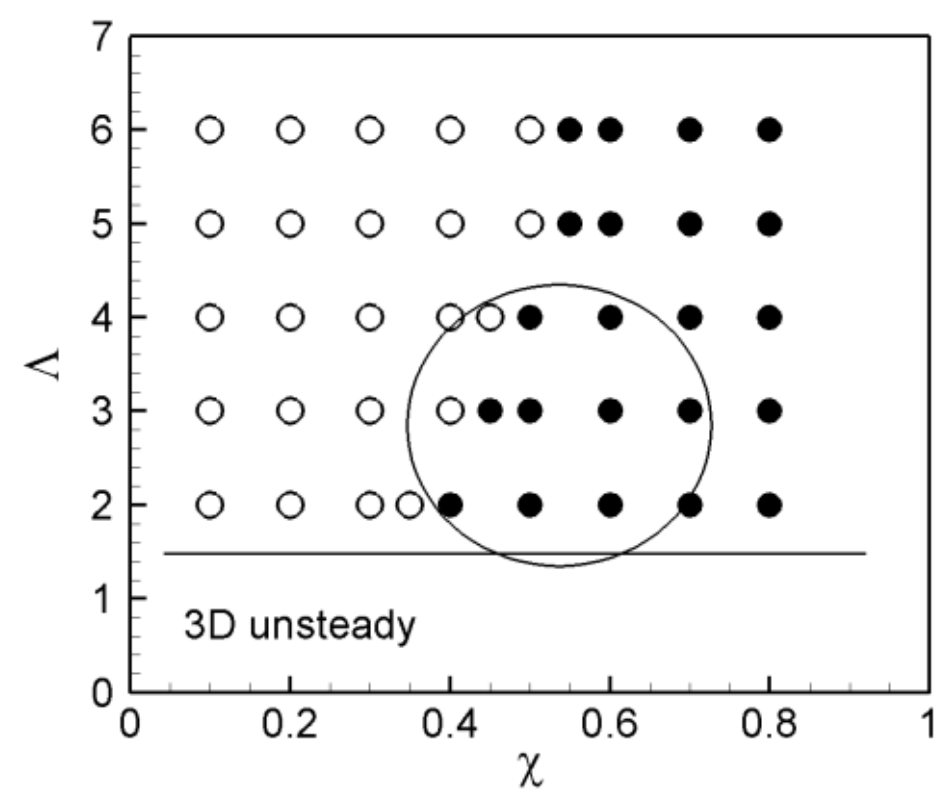

FIG. 12. Regime diagram of bolus formation for the moderate-inertia case ($Re = 100$) over the parameter range $2 \leq \Lambda \leq 6$ and $0.1 < \chi < 0.8$. Filled symbols denote cases exhibiting bolus formation, while open symbols indicate flows without trapping. The $\Lambda = 1$ cases are not considered further because the flow becomes three-dimensional and unsteady. The circled region identifies a short-wavelength trapping regime that emerges at relatively small wave amplitudes.

Figure 12 presents the bolus-formation regime diagram for the moderate-inertia case, $Re = 100$. The $\Lambda = 1$ cases are not considered further because the flow becomes three-dimensional and unsteady. For the longest wavelengths ($\Lambda = 6$ and 5), the critical wave amplitude required for bolus formation increases compared with the $Re = 20$ case, reaching approximately $0.5 < \chi < 0.55$. This shift suggests that increasing inertia strengthens the interaction between the throat flow and the trapped bolus, making bolus formation increasingly difficult.

In contrast, as the wavelength decreases from $\Lambda = 4$ to 2, the critical wave amplitude for bolus formation decreases progressively with decreasing $\Lambda$. This trend is characteristic of the curvature-induced trapping mechanism identified in the weak-inertia regime. Compared with the $Re = 20$ results, the curvature-induced trapping regime extends from the shortest-wavelength limit to a broader range of wavelengths. These results demonstrate that increasing inertia enlarges the region of parameter space over which the curvature-induced trapping mechanism becomes important, extending its influence toward longer wavelengths.

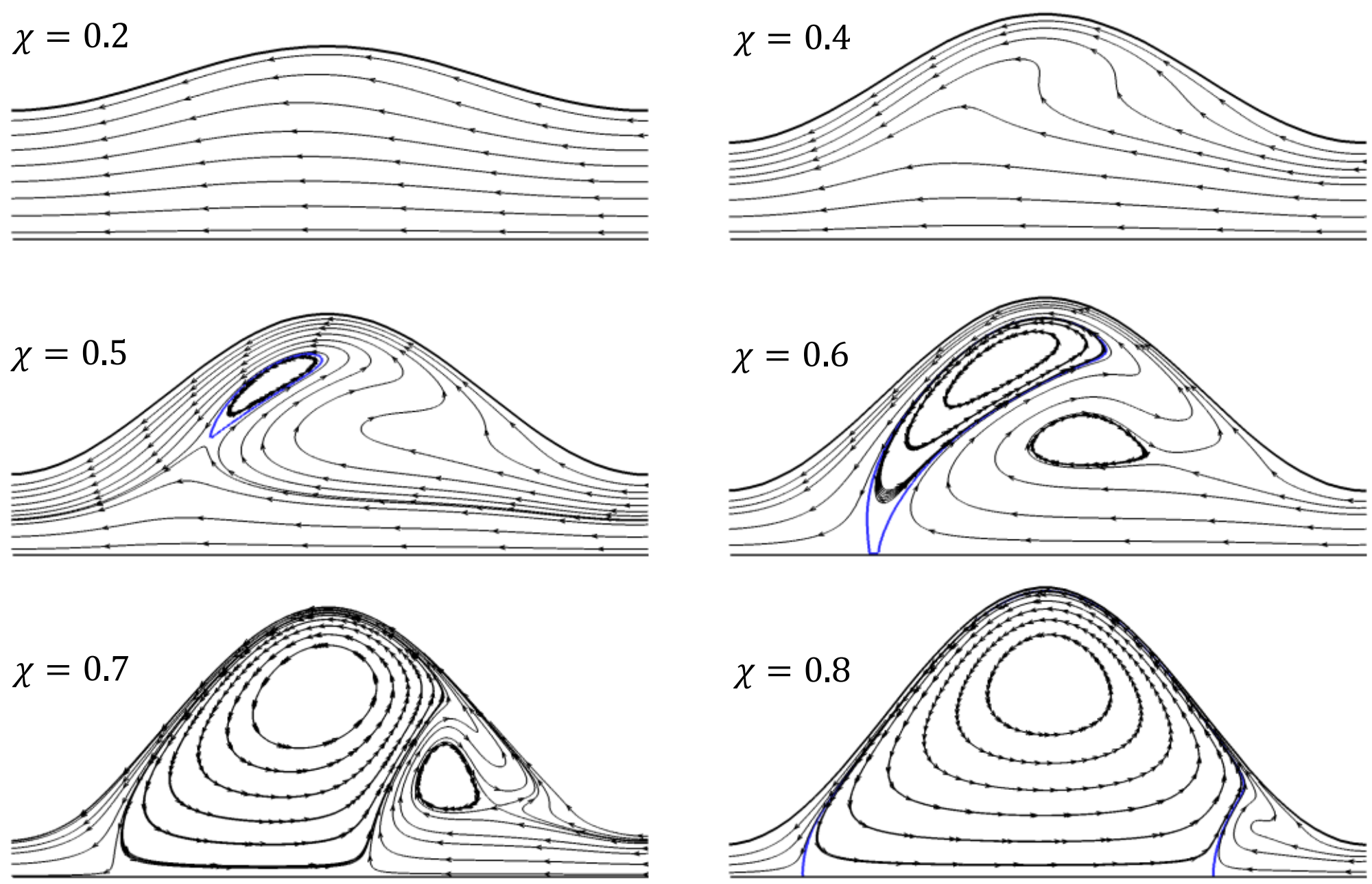

FIG. 13. Streamline patterns in the wave frame at $Re = 100$ and $\Lambda = 4$ for wave amplitude ratios $0.2 \leq \chi \leq 0.8$. Blue contours indicate the bolus boundaries. No trapping is observed at $\chi = 0.2$ and 0.4. At $\chi = 0.5$, trapping first appears as a small off-axis bolus, indicating the dominance of the curvature-induced trapping mechanism. As the wave amplitude increases, the primary bolus grows in size and migrates toward the centerline. Additional boluses appear on the right side of the primary structure at $\chi = 0.6$ and 0.7; these secondary recirculating regions arise from the deflection of the throat flow after its interaction with the primary bolus and are not associated with the two principal trapping mechanisms discussed in this work. By $\chi = 0.7$, the primary bolus reaches the centerline and the classical trapping mechanism becomes dominant. Further increasing the wave amplitude strengthens the centerline bolus, which occupies most of the wave interior at $\chi = 0.8$, illustrating the transition from curvature-induced trapping to classical trapping.

To further elucidate the interaction between the classical trapping mechanism and the curvature-induced trapping mechanism, we examine two representative parameter paths within the regime diagram of Fig. 12. Specifically, Fig. 13 fixes the wavelength ratio at $\Lambda = 4$ and follows the evolution of the flow structure with increasing wave amplitude, while Fig. 14 fixes the wave amplitude ratio at $\chi = 0.6$ and illustrates the effect of decreasing wavelength.

Figure 13 demonstrates how the dominant trapping mechanism changes with wave amplitude. At $\chi = 0.5$, a small off-axis bolus first appears near the wave crest. Its location and elongated shape are characteristic of the curvature-induced trapping mechanism. Similar flow structures were observed by Xiao and Damodaran at comparable Reynolds numbers [15]. However, the physical mechanisms responsible for their formation were not investigated further. As the wave amplitude increases to $\chi = 0.6$, the bolus grows rapidly, indicating that the curvature-induced trapping mechanism becomes increasingly important. At the same time, a secondary recirculating region appears on the right side of the primary bolus. This structure is generated by the deflection of the throat flow after its interaction with the primary bolus and should not be regarded as an independent trapping mechanism. As the wave amplitude increases further to $\chi = 0.7$, the primary bolus expands toward the centerline and the classical trapping structure becomes dominant. At $\chi = 0.8$, the centerline bolus occupies most of the wave interior, indicating that the classical trapping mechanism has become the primary contributor to bolus formation. These results reveal a continuous transition from curvature-induced trapping at moderate wave amplitudes to classical trapping at larger wave amplitudes.

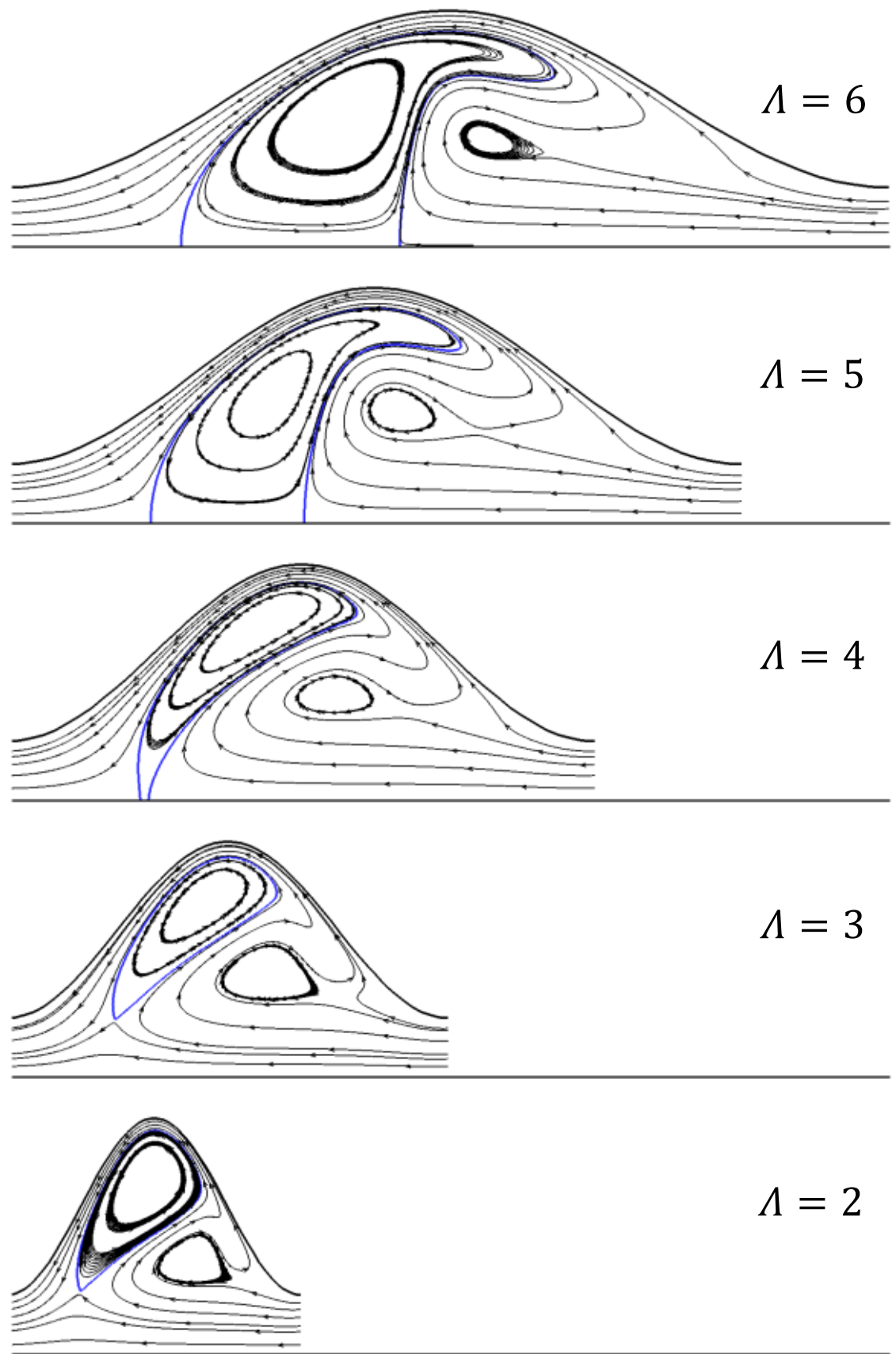

FIG. 14. Streamline patterns in the wave frame at $Re = 100$ and $\chi = 0.6$ for wavelength ratios ($2 \le \Lambda \le 6$). Blue contours indicate the bolus boundaries. For the longer wavelengths ($\Lambda = 6$ and 5), the primary bolus remains largely centerline-centered, characteristic of the classical trapping mode, although noticeable bending of the streamlines near the wave crest indicates the influence of wave curvature. As the wavelength decreases, curvature effects become increasingly pronounced, causing the primary bolus to migrate away from the centerline and adopt a more elongated shape aligned with the curved wall. For $\Lambda = 3$ and 2, the trapping structure is dominated by the off-axis bolus associated with the curvature-induced trapping mechanism, illustrating the progressive transition from classical trapping to curvature-dominated trapping as the wavelength decreases.

The complementary evolution obtained by varying wavelength is shown in Fig. 14. For the longest wavelengths ($\Lambda = 6$ and 5), the primary bolus remains largely centered on the axis, consistent with the classical trapping mechanism. Nevertheless, the noticeable bending of streamlines near the wave crest indicates that curvature effects are already present. As the wavelength decreases, the influence of wave curvature becomes progressively stronger, causing the bolus to migrate away from the centerline and align with the curved wall. For $\Lambda = 3$ and 2, the off-axis bolus becomes the dominant trapping structure, indicating that the curvature-induced trapping mechanism has overtaken the classical mechanism. Together, Figs. 13 and 14 demonstrate that increasing wave amplitude favors the classical trapping pathway, whereas decreasing wavelength promotes the curvature-induced pathway. The trapping behavior observed at $Re = 100$ therefore results from the competition between these two mechanisms, whose relative importance depends strongly on both wavelength and wave amplitude.

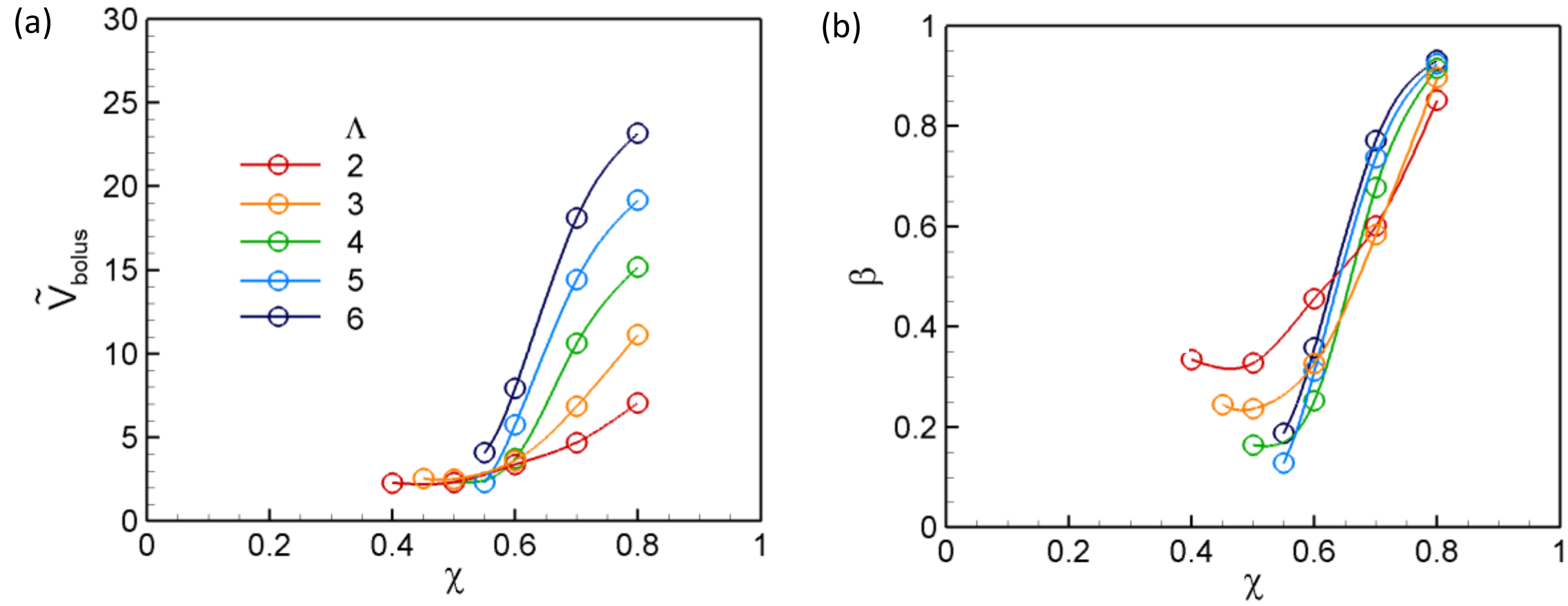

FIG. 15. Variation of the dimensionless bolus volume $\tilde{V}_{bolus}$ and bolus volume fraction $\beta$ with wave amplitude $\chi$ at $Re = 100$ for different wavelength ratios $\Lambda$. (a) Dimensionless bolus volume $\tilde{V}_{bolus}$ as a function of $\chi$. (b) Bolus volume fraction $\beta$ as a function of $\chi$. The short-wavelength cases exhibit trends distinct from those of the longer wavelengths, reflecting the increasingly important role of wavelength-dependent trapping dynamics at moderate inertia.

Figure 15 provides a quantitative view of the trapping dynamics at $Re = 100$ through the evolution of the bolus volume and bolus volume fraction. Consistent with the streamline topologies shown in Figs. 13 and 14, the evolution curves gradually change their character as the wavelength decreases, reflecting the growing influence of the curvature-induced trapping mechanism. The most notable feature is observed near the onset of trapping. For all wavelengths considered, both the bolus volume and bolus volume fraction exhibit a similar upward-bending trend as the wave amplitude approaches the trapping threshold. Such behavior was previously observed only in the short-wavelength regime at lower Reynolds numbers. Its appearance across the entire wavelength range at $Re = 100$ indicates that the influence of the curvature-induced trapping mechanism is no longer confined to the shortest wavelengths. As the wave amplitude increases, the influence of the classical trapping mechanism becomes progressively stronger. Consequently, the curvature-induced signatures that dominate near the trapping threshold become less pronounced, and the subsequent evolution of the bolus is increasingly governed by the classical trapping dynamics. These observations, together with the regime diagram in Fig. 12 and the streamline topologies in Figs. 13 and 14, demonstrate that increasing inertia not only strengthens the curvature-induced trapping mechanism, but also extends its influence over a much broader region of the parameter space.

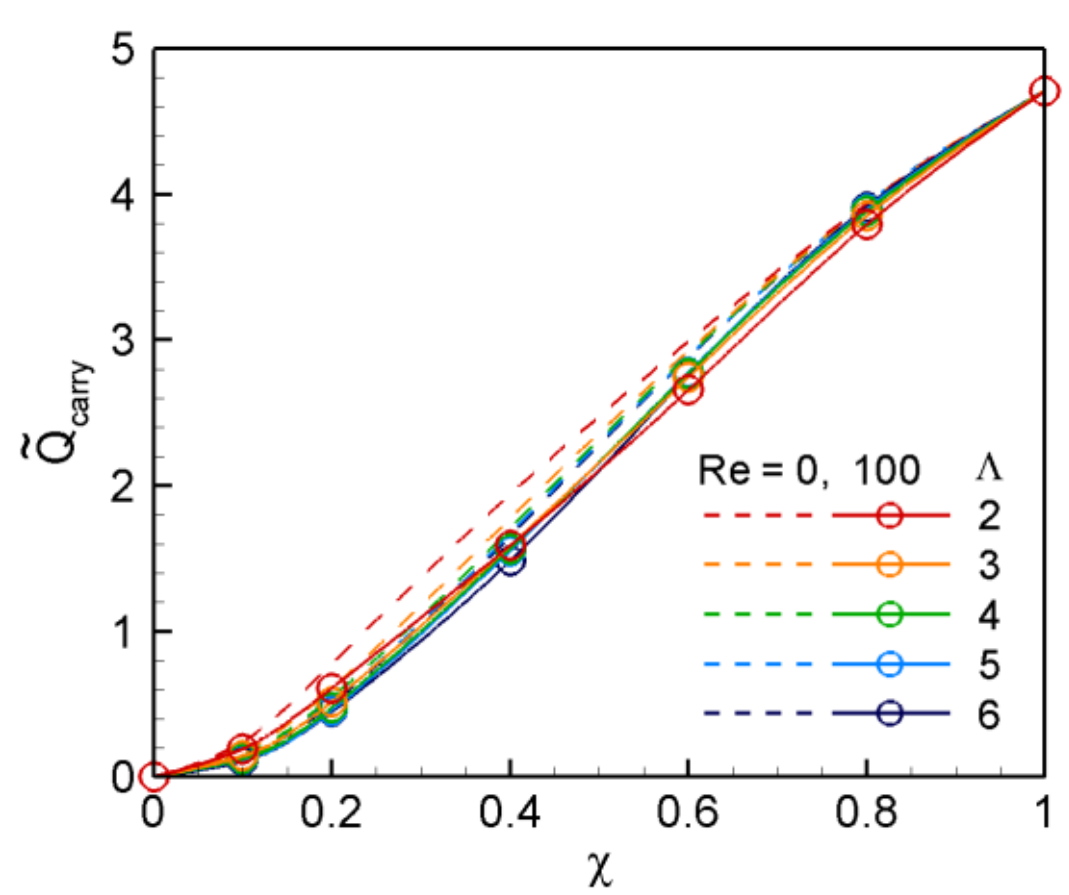

FIG. 16. Variation of the dimensionless volumetric flow rate $\tilde{Q}_{carry}$ with wave amplitude $\chi$ for different wavelength ratios $\Lambda$ at $Re = 100$. The corresponding Stokes-flow results ($Re = 0$, dashed lines) are included for comparison. Although noticeable deviations from the Stokes-flow behavior emerge at intermediate wave amplitudes, particularly for shorter wavelengths, the overall transport rate remains only weakly affected by inertia.

Figure 16 compares the dimensionless volumetric flow rates at $Re = 100$ with those obtained in the Stokes limit. Despite the substantial changes observed in the trapping topology, bolus evolution, and the expansion of the curvature-induced trapping regime, the overall transport rate remains relatively insensitive to inertia. For all wavelengths considered, the $Re = 100$ results remain close to their Stokes-flow counterparts, indicating that inertia has a much stronger influence on trapping dynamics than on the net transport capacity of the flow. The largest deviations from the Stokes-flow results occur at intermediate wave amplitudes, where the competition between the classical and curvature-induced trapping mechanisms is most pronounced. However, as the wave amplitude increases further, the transport-rate curves corresponding to different Reynolds numbers gradually converge. These observations suggest that the primary effect of inertia is to reorganize the internal flow structure and trapping topology, while the net volumetric transport remains largely controlled by the wave amplitude itself.

To gain further insight into the mechanism of curvature-induced trapping, a simple scaling argument is proposed. It should be emphasized that the term curvature-induced trapping is adopted here to reflect the pronounced geometric bending associated with short-wavelength waves rather than the mathematical curvature of the wall. The mechanism originates from the turning of fluid trajectories induced by the wave geometry near the wave crest. This turning effect may be characterized by a geometric turning parameter,

$$\Theta_t = \frac{\chi}{\Lambda} \tag{5.8}$$

where larger values of $\Theta_t$ correspond to stronger geometric turning and therefore a greater tendency for local recirculating motion to develop within the wave-crest region. Geometric turning alone, however, does not guarantee the formation of a trapped bolus. The resulting recirculating motion must persist against viscous diffusion within the wave-crest region. Since the characteristic size of this region scales with the wave amplitude $A$, a pocket Reynolds number may be defined as

$$Re_p = \frac{c_w A}{\nu} = \chi Re \tag{5.9}$$

which characterizes the relative importance of inertial transport and viscous diffusion within the wave-crest pocket. The tendency for curvature-induced trapping may therefore be represented in the general form

$$\Pi_c = F\left(\Theta_t, Re_p\right) \tag{5.10}$$

where $F$ is an increasing function of both arguments. Physically, this implies that curvature-induced trapping becomes more favorable as the geometric turning becomes stronger and as the resulting recirculating motion becomes more resistant to viscous diffusion. As a first-order approximation, the dependence may be expressed in separable form,

$$\Pi_c = F_1(\Theta_t) F_2\left(Re_p\right) \tag{5.11}$$

where both $F_1$ and $F_2$ are monotonically increasing functions. Taking the simplest linear approximation,

$$F_1(\Theta_t) \sim \Theta_t \tag{5.12}$$

and

$$F_2\left(Re_p\right) \sim Re_p \tag{5.13}$$

gives

$$\Pi_c \sim \Theta_t Re_p = \frac{\chi^2}{\Lambda} Re \tag{5.14}$$

Therefore, curvature-induced trapping is expected to become increasingly favorable as the combined parameter $\frac{\chi^2}{\Lambda} Re$ increases. This trend is also consistent with the regime-diagram results presented in Fig. 12, where curvature-induced trapping becomes increasingly favorable with decreasing wavelength and increasing Reynolds number. It should be emphasized, however, that the proposed scaling is intended only as a first-order estimate of the dominant trend rather than a quantitative prediction of the trapping boundary. The interaction with the centerline through-flow, together with the coexistence and competition of the classical and curvature-induced trapping mechanisms, can modify the precise onset and extent of the trapping regime. Consequently, the proposed scaling should be interpreted as an organizing framework for understanding the observed parameter dependence rather than as a universal trapping criterion.

### D. Inertial evolution and interaction of trapping pathways

Figure 17 summarizes the influence of inertia on bolus morphology through the evolution of the bolus volume fraction $\beta$. Solid lines denote regimes dominated by the classical trapping mechanism, whereas dashed lines denote regimes dominated by the curvature-induced trapping mechanism.

For $\chi = 0.8$, the bolus volume fraction decreases monotonically with increasing Reynolds number for all wavelengths considered. The reduction becomes more pronounced as the wavelength decreases. This behavior results from the enhanced impact of the throat flow on the trapped bolus at higher Reynolds numbers, which progressively suppresses the classical trapping structure. For $\chi = 0.6$, the evolution begins to depend strongly on wavelength. The $\Lambda = 6$ case continues to exhibit a monotonic decrease in $\beta$, indicating that the bolus evolution remains governed primarily by the classical trapping mechanism. The $\Lambda = 4$ case develops a slight recovery at larger Reynolds numbers, suggesting that wave-curvature effects have begun to influence the trapping dynamics, although the classical trapping mechanism remains dominant. As the wavelength decreases further to $\Lambda = 2$, the curve separates into two stages: $\beta$ first decreases with Reynolds number as the classical trapping structure is weakened, and then increases after the curvature-induced trapping mechanism becomes dominant. This decrease-and-growth behavior directly reflects the transition between the two trapping pathways. For $\chi = 0.4$, a similar progression is observed, but the transition occurs at lower Reynolds numbers and the curvature-induced regime occupies a larger portion of the Reynolds-number range. Consequently, the increase in $\beta$ associated with the curvature-induced bolus becomes more pronounced.

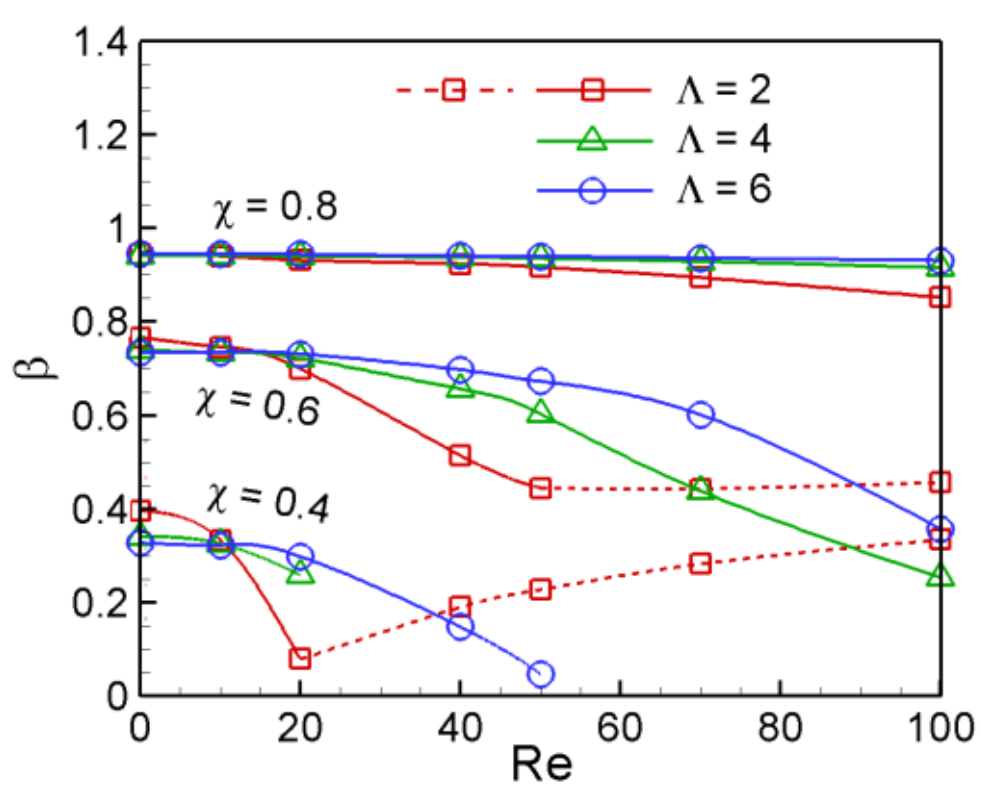

FIG. 17. Variation of the bolus volume fraction $\beta$ with Reynolds number $Re$ for representative wavelength ratios ($\Lambda = 2$, 4 and 6) and wave amplitude ratios ($\chi = 0.4$, 0.6, and 0.8). Solid lines denote regimes in which the classical trapping mechanism is dominant, whereas dashed lines denote regimes dominated by the curvature-induced trapping mechanism.

Taken together, these results reveal a continuous transition from a classical-trapping-dominated regime to a curvature-influenced regime and eventually to a curvature-induced-trapping-dominated regime. Increasing inertia therefore plays a dual role: it suppresses the classical bolus through the enhanced impact of the throat flow while simultaneously promoting the development of curvature-induced trapping. The observed bolus morphology is ultimately determined by the competition between these two effects, whose relative importance depends on Reynolds number, wavelength, and wave amplitude.

## VI. CONCLUSION

In this work, we investigated bolus formation in axisymmetric peristaltic-wave-driven flows under zero-pressure-gradient conditions using a high-fidelity lattice Boltzmann framework. By systematically varying wavelength, wave amplitude, and Reynolds number from the Stokes limit to moderate inertia, we established a mechanistic picture of how flow organization and trapping evolve in peristaltic-wave-driven transport.

In the Stokes limit, bolus formation is governed by the classical trapping mechanism, in which a centerline recirculating bolus develops beneath the wave crest. Despite the large variation in wavelength, the resulting flow structures exhibit a remarkable degree of similarity. When the axial coordinate is normalized by the wavelength, both the bolus morphology and transport characteristics progressively approach common asymptotic behaviors as the wavelength increases. This wavelength-normalized similarity indicates that long-wavelength peristaltic flows share a common underlying flow organization.

The introduction of inertia progressively breaks this similarity. For sufficiently long wavelengths, the overall trapping behavior remains qualitatively similar to that observed in the Stokes limit, although the trapped bolus is increasingly suppressed by the interaction between the throat flow and the recirculating region. As a result, the wave-amplitude threshold required for bolus formation increases with Reynolds number.

A fundamentally different behavior emerges in the short-wavelength regime. As the wavelength decreases, the strongly curved wave geometry reorganizes the local flow structure and produces an off-axis bolus that is distinct from the classical centerline trapping mode. This curvature-induced trapping mechanism first appears in the shortest-wavelength cases at weak inertia and expands progressively toward longer wavelengths as Reynolds number increases. At moderate inertia, wave-curvature effects influence a substantial portion of the parameter space and become an important organizing factor governing bolus formation.

The interaction between these two trapping pathways determines the observed bolus morphology. Increasing wave amplitude promotes the classical trapping mechanism, whereas decreasing wavelength favors the curvature-induced pathway. Reynolds number plays a dual role by simultaneously suppressing the classical bolus and promoting curvature-induced trapping. As a result, the flow evolves continuously from a classical-trapping-dominated regime to a curvature-influenced regime and ultimately to a curvature-induced-trapping-dominated regime.

Overall, the present study demonstrates that bolus formation in peristaltic-wave-driven flows cannot be described by a single universal trapping mechanism. Instead, it is governed by the interplay among wavelength-dependent flow similarity, wave-curvature effects, and fluid inertia. These findings provide a unified physical framework for understanding trapping and transport in travelling-wave-driven flows and offer new insight into peristaltic transport processes occurring in both physiological and engineering systems.

**Acknowledgements:** This research was supported by the U.S. National Science Foundation (NSF) through the CAREER Program under Award No. 2443848.

[1] J. C. Burns and T. Parkes, Peristaltic motion, J. Fluid Mech. 29, 731 (1967).
[2] M. Hanin, The flow through a channel due to transversely oscillating walls, Isr. J. Technol. 6, 67 (1968).
[3] Y. C. Fung and C. S. Yih, Peristaltic transport, J. Appl. Mech. 35, 669 (1968).
[4] A. H. Shapiro, M. Y. Jaffrin, and S. L. Weinberg, Peristaltic pumping with long wavelengths at low Reynolds number, J. Fluid Mech. 37, 799 (1969).
[5] F. Yin and Y. C. Fung, Peristaltic waves in circular cylindrical tubes, J. Appl. Mech. 36, 579 (1969).
[6] T. W. Latham, Fluid motions in a peristaltic pump, S.M. thesis, Massachusetts Institute of Technology, Cambridge, MA, 1966.
[7] S. L. Weinberg, A theoretical and experimental treatment of peristaltic pumping and its relation to ureteral function, Ph.D. thesis, Massachusetts Institute of Technology, Cambridge, MA, 1970.
[8] E. C. Eckstein, Experimental and theoretical pressure studies of peristaltic pumping, S.M. thesis, Massachusetts Institute of Technology, Cambridge, MA, 1970.
[9] M. Y. Jaffrin and A. H. Shapiro, Peristaltic pumping, Annu. Rev. Fluid Mech. 3, 13 (1971).
[10] M. Y. Jaffrin, Inertia and streamline curvature effects on peristaltic pumping, Int. J. Eng. Sci. 11, 681 (1973).
[11] T. D. Brown and T. K. Hung, Computational and experimental investigations of two-dimensional nonlinear peristaltic flows, J. Fluid Mech. 83, 249 (1977).
[12] S. Takabatake and K. Ayukawa, Numerical study of two-dimensional peristaltic flows, J. Fluid Mech. 122, 439 (1982).
[13] C. Pozrikidis, A study of peristaltic flow, J. Fluid Mech. 180, 515 (1987).
[14] S. Takabatake, K. Ayukawa, and A. Mori, Peristaltic pumping in circular cylindrical tubes: A numerical study of fluid transport and its efficiency, J. Fluid Mech. 193, 267 (1988).
[15] Q. Xiao and M. Damodaran, A numerical investigation of peristaltic waves in circular tubes, Int. J. Comput. Fluid Dyn. 16, 201 (2002).
[16] M. Li and J. G. Brasseur, Non-steady peristaltic transport in finite-length tubes, J. Fluid Mech. 248, 129 (1993).
[17] A. R. Rao and S. Usha, Peristaltic transport of two immiscible viscous fluids in a circular tube, J. Fluid Mech. 298, 271 (1995).
[18] M. Mishra and A. Ramachandra Rao, Peristaltic transport of a Newtonian fluid in an asymmetric channel, Z. Angew. Math. Phys. 54, 532 (2003).
[19] P. Hariharan, V. Seshadri, and R. K. Banerjee, Peristaltic transport of non-Newtonian fluid in a diverging tube with different wave forms, Math. Comput. Modell. 48, 998 (2008).
[20] D. Takagi and N. J. Balmforth, Peristaltic pumping of viscous fluid in an elastic tube, J. Fluid Mech. 672, 196 (2011).
[21] S. Kumar, H. J. Kim, and A. Beskok, Numerical simulations of peristaltic mixing, J. Fluids Eng. 129, 1361 (2007).
[22] D. Tripathi and O. A. Bég, Transient magneto-peristaltic flow of couple stress biofluids: A magneto-hydro-dynamical study on digestive transport phenomena, Math. Biosci. 246, 72 (2013).
[23] J. Cremer, I. Segota, C. Y. Yang, M. Arnoldini, J. T. Sauls, Z. Zhang, E. Gutierrez, A. Groisman, and T. Hwa, Effect of flow and peristaltic mixing on bacterial growth in a gut-like channel, Proc. Natl. Acad. Sci. U.S.A. 113, 11414 (2016).

[24] R. Ibanez, M. Shokrian, J.-H. Nam, and D. H. Kelley, Simple analytic model for peristaltic flow and mixing, Phys. Rev. Fluids 6, 103101 (2021).
[25] R. J. Amedzrovi Agbesi and N. R. Chevalier, Flow and mixing induced by single, colinear, and colliding contractile waves in the intestine, Phys. Rev. Fluids 7, 043101 (2022).
[26] R. Ebara, S. Ishida, T. Miyagawa, and Y. Imai, Effects of peristaltic amplitude and frequency on gastric emptying and mixing: A simulation study, J. R. Soc. Interface 20, 20220780 (2023).
[27] Y. Qian, D. d'Humières, and P. Lallemand, Lattice BGK models for Navier-Stokes equation, Europhys. Lett. 17, 479 (1992).
[28] S. Chen and G. D. Doolen, Lattice Boltzmann method for fluid flows, Annu. Rev. Fluid Mech. 30, 329 (1998).
[29] Y. Wang, J. G. Brasseur, G. G. Banco, A. G. Webb, A. C. Ailiani, and T. Neuberger, Development of a lattice-Boltzmann method for multiscale transport and absorption with application to intestinal function, in Computational Modeling in Biomechanics, edited by S. De, F. Guilak, and M. Mofrad (Springer, Dordrecht, 2010), pp. 327–354.
[30] P. L. Bhatnagar, E. P. Gross, and M. Krook, A model for collision processes in gases. I. Small amplitude processes in charged and neutral one-component systems, Phys. Rev. 94, 511 (1954).
[31] P. Lallemand and L.-S. Luo, Lattice Boltzmann method for moving boundaries, J. Comput. Phys. 184, 406 (2003).
[32] A. J. C. Ladd, Numerical simulations of particulate suspensions via a discretized Boltzmann equation. Part 1. Theoretical foundation, J. Fluid Mech. 271, 285 (1994).
[33] M. Bouzidi, M. Firdaouss, and P. Lallemand, Momentum transfer of a Boltzmann-lattice fluid with boundaries, Phys. Fluids 13, 3452 (2001).
[34] B. Chun, and A.J. Ladd, Interpolated boundary condition for lattice Boltzmann simulations of flows in narrow gaps. Physical Review E—Statistical, Nonlinear, and Soft Matter Physics, 75, 066705 (2007).
[35] Y. Wang, H. Wan, T. Wei, and F. Shu, Fluid dynamics and passive scalar transport driven by non-uniform tumbling of a prolate spheroid in simple shear flow, J. Fluid Mech. 1021, A11 (2025).
[36] Y. Wang, H. Wan, R. Gonzalez Pizarro, S. Lim, and F. Shu, Heat and mass transport from a neutrally suspended oblate spheroid in simple shear flow, Phys. Fluids 35, 033319 (2023).
[37] Y. Wang, J. G. Brasseur, G. G. Banco, A. G. Webb, A. C. Ailiani, and T. Neuberger, A multiscale lattice Boltzmann model of macro- to micro-scale transport, with applications to gut function, Philos. Trans. R. Soc. A 368, 3863 (2010).
[38] G.V. Levina, Some regimes of peristaltic pumping, Fluid Dynamics, 18, 683 (1983).
[39] G.V. Levina, Peristaltic flow at finite Reynolds numbers, Fluid Dynamics, 20, 344 (1985).